\pdfoutput=1
\documentclass[11pt,twoside,a4paper,cmspaper,final,collab]{cms-tdr}
\def\svnVersion{f520e53-D}\def\svnDate{2021/05/20}\def\cmsCernNoTag{CERN-EP-2020-216}\def\cmsCernDate{\today}\def\cmsMessage{"Published in Physics Letters B as \href{http://dx.doi.org/10.1016/j.physletb.2021.136446}{\doi{10.1016/j.physletb.2021.136446}.}"}
\begin{document}\cmsNoteHeader{EXO-19-015}

\newcommand{\runtwolumi}{137\xspace}  
\newcommand{\LQ}{\ensuremath{\text{LQ}}\xspace}
\newcommand{\LQs}{\ensuremath{\text{leptoquarks}}\xspace}
\newcommand{\ALQ}{\ensuremath{\overline{\LQ}}\xspace}
\newcommand{\LQS}{\ensuremath{\text{LQ}_{\text{S}}}\xspace}
\newcommand{\LQV}{\ensuremath{\text{LQ}_{\text{V}}}\xspace}
\newcommand{\ALQS}{\ensuremath{\overline{\LQ}_{\text{S}}}\xspace}
\newcommand{\ALQV}{\ensuremath{\overline{\LQ}_{\text{V}}}\xspace}
\newcommand{\ST}{\ensuremath{S_{\mathrm{T}}}\xspace}
\newcommand{\nbjets}{\ensuremath{N_{{\PQb}\text{-jet}}}\xspace}
\newlength\cmsTabSkip\setlength{\cmsTabSkip}{1ex}

\cmsNoteHeader{EXO-19-015} 
\title{Search for singly and pair-produced leptoquarks coupling to third-generation fermions in proton-proton collisions at \texorpdfstring{$\sqrt{s} = 13\TeV$}{sqrt(s)=13 TeV}}

\date{\today}

\abstract{
A search for leptoquarks produced singly and in pairs in proton-proton collisions is presented.
We consider the leptoquark (\LQ) to be a scalar particle of charge -1/3$e$ coupling to a top quark plus a tau lepton ($\PQt\PGt$) or a bottom quark plus a neutrino ($\PQb\PGn$), or a vector particle of charge +2/3$e$, coupling to $\PQt\PGn$ or $\PQb\PGt$. These choices are motivated by models that can explain a series of anomalies observed in the measurement of \PB meson decays.  
In this analysis the signatures $\PQt\PGt\PGn\PQb$ and $\PQt\PGt\PGn$ are probed, 
using data recorded by the CMS experiment at the CERN LHC at $\sqrt{s} = 13\TeV$ and that correspond to an integrated luminosity of $\runtwolumi\fbinv$.
These signatures have not been previously explored in a dedicated search.
The data are found to be in agreement with the standard model prediction. 
Lower limits at 95\% confidence level are set on the \LQ mass in the range 0.98--1.73\TeV, depending on the \LQ spin and its coupling $\lambda$ to a lepton and a quark, and assuming equal couplings for the two \LQ decay modes considered.
These are the most stringent constraints to date on the existence of leptoquarks in this scenario.
}

\hypersetup{
pdfauthor={CMS Collaboration},
pdftitle={Search for singly- and pair-produced leptoquarks coupled to third-generation fermions in proton-proton collisions at sqrt{s} = 13 TeV},
pdfsubject={CMS},
pdfkeywords={CMS, search for new physics, leptoquarks}}

\maketitle 
\section{Introduction}
\label{sec:Introduction}
Experimental evidence has promoted the standard model (SM) to the role of a reference theory of the physics of elementary particles.
Despite the theory's successes, there are several fundamental aspects of observed particle physics that lack a complete explanation. 
One of these is the symmetry between the quark and lepton families.
Possible explanations have been offered by several models that extend the SM, such as grand unified theories~\cite{Pati:1973uk, PatiSalam, GeorgiGlashow, Fritzsch:1974nn},
technicolor models~\cite{Dimopoulos:1979es,Dimopoulos:1979sp,Technicolor, Lane:1991qh},
compositeness scenarios~\cite{LightLeptoquarks,Gripaios:2009dq},
and $R$-parity violating supersymmetry~\cite{Farrar:1978xj,Ramond:1971gb,Golfand:1971iw,Neveu:1971rx,Volkov:1972jx,Wess:1973kz,Wess:1974tw,Fayet:1974pd,Nilles:1983ge,Barbier:2004ez}.
These theories foresee a new particle that carries both lepton number $L$ and baryon number $B$, and is generically referred to as a ``leptoquark'' (\LQ). 

\begin{figure*}[t]
\centering
\includegraphics[width=0.9\textwidth]{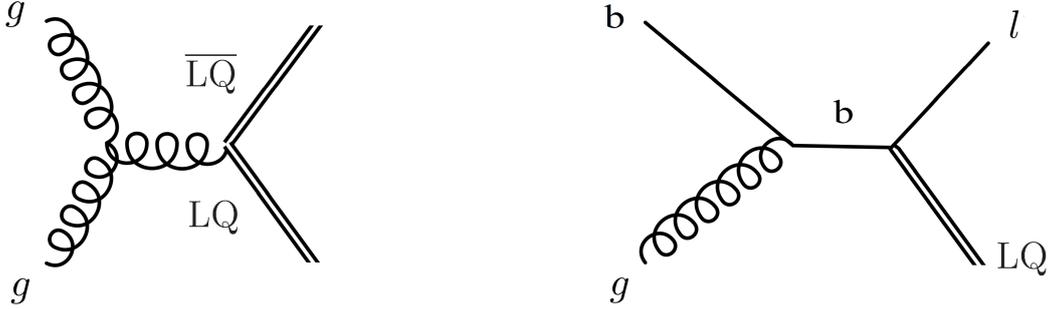}
\caption{
Feynman diagrams for dominant leptoquark production modes at leading order: pairwise (left), and in combination with a lepton (right). 
In the scenarios considered the \LQS may couple to $\PQt\PGt$ or $\PQb\PGn$, while the \LQV may couple to $\PQt\PGn$ or $\PQb\PGt$.
}
\label{fig:LQtopFeynmanDiagram}
\end{figure*}

A leptoquark has a fractional electric charge, and can be either a scalar particle (\LQS, with a spin of 0), or a vector particle (\LQV, with a spin of 1), with 3$B+L$ equal to either 2 or 0.
At hadron colliders, \LQs can be produced in pairs, or singly in association with a lepton~\cite{Diaz:2017lit,Schmaltz:2018nls}, as illustrated by the Feynman diagrams in Fig.~\ref{fig:LQtopFeynmanDiagram}. 
For \LQS pair production, the cross section depends only on the \LQS mass for the range of \LQ mass and $\lambda$ values investigated in this search, while for \LQV it may depend on additional parameters~\cite{Blumlein:1996qp}, to comply with constraints imposed by unitarity at high energy scales. For singly produced \LQs, 
the cross section further depends on the couplings of the \LQ to the quark and the lepton, and on the quark flavor.

Leptoquarks have recently gained enhanced interest, as they may provide an explanation for a series of anomalies observed in the measurement of \PB\ meson decays in 
charged-current 
$\PQb\to\PQc\ell\PGn$~\cite{Lees:2012xj,Lees:2013uzd,Huschle:2015rga,Sato:2016svk,Hirose:2016wfn,Hirose:2017dxl, Belle:2019rba,
Aaij:2015yra,Aaij:2017uff,Aaij:2017tyk} and
neutral-current 
$\PQb\to\PQs\ell\ell$~\cite{Wehle:2016yoi,Aaij:2013qta,Aaij:2014pli,Aaij:2014ora,Aaij:2015oid,Aaij:2015esa,Aaij:2017vbb,Aaij:2019wad} processes.
The solutions proposed to explain these anomalies favor effective couplings to third-generation SM fermions at the\TeV scale, leading to processes that may be accessible at the CERN LHC. 
In particular, the model of Ref.~\cite{Alvarez:2018gxs} predicts a charge -1/3$e$ \LQS, with 3$B+L = 2$, decaying to a top quark and a $\tau$ lepton ($\PQt\PGt$), or a
bottom quark and a neutrino ($\PQb\PGnGt$), while the model presented in Ref.~\cite{Buttazzo:2017ixm} contains a charge +2/3$e$ \LQV, with 3$B+L = 0$, decaying to 
a top quark and an antineutrino ($\PQt\PAGnGt$) or a bottom quark and an anti-$\tau$ lepton ($\PQb\PGtp$).
Each model includes a charge-conjugate leptoquark and prefers a region of parameter space that gives equal branching fractions for the two allowed decays, rendering the $\PQt\PGt\PGn\PQb$ signature as the most frequent for pair-produced leptoquarks.

The analysis described in this Letter investigates the
existence of \LQs produced in pairs with decays leading to the $\PQt\PQb\PGt\PGn$ signature, or singly with the decay leading to $\PQt\PGt\PGn$.
The models of Refs.~\cite{Alvarez:2018gxs,Buttazzo:2017ixm} are considered in this analysis, relying on the implementations described in Refs.~\cite{Buchmuller1987442,Dorsner:2018ynv}.
In these models, the parameters of interest for determining the cross section are: 
the \LQ mass; 
for \LQV, a dimensionless coupling $k$, set to 1 (Yang--Mills case) or 0 (minimal coupling case)~\cite{Blumlein:1996qp}; 
and the \LQ coupling ($\lambda$) to the lepton and quark, which affects the cross section for single \LQ production.  We note that the analysis is designed to be agnostic to the charge of the \LQ, and is thus sensitive also to models with up-type scalar \LQ and down-type vector \LQ, which are not directly considered below. 

The most recent searches for leptoquarks have been performed at $\sqrt{s} = 13\TeV$ by the ATLAS and CMS Collaborations for couplings to ($\PQt\PGt$, $\PQb\PGn$) and ($\PQt\PGn$, $\PQb\PGt$)~\cite{ATLAS:2013oea,Aad:2015caa,Aaboud:2019bye,Sirunyan:2018kzh,Sirunyan:2018nkj,Sirunyan:2018ruf,Sirunyan:2018vhk,Sirunyan:2018jdk} and for couplings to other quark-lepton pairs~\cite{Aad:2020iuy, Aaboud:2019jcc, Aaboud:2016qeg,Sirunyan:2018ryt, Sirunyan:2018ruf}.

Differently from previous searches that have separately
considered single or pair \LQ production, the present analysis
strategy is devised to search for both production mechanisms
simultaneously. The $\PQt\PGt\PGn$($\PQb$) signatures are analyzed for the first time considering the inclusive hadronic decay channels of the top quark and $\PGt$ lepton. We include a dedicated selection for the case of a large LQ-t mass splitting giving rise to a Lorentz-boosted top quark, whose decay products may not be resolved as individual jets.

The search is based on a data sample of proton-proton ($\Pp\Pp$) collisions at a center-of-mass energy of 13\TeV recorded by the CMS experiment at the CERN LHC in the years 2016--18, corresponding to an integrated luminosity of $\runtwolumi\fbinv$.

\section{The CMS detector}
\label{sec:The CMS detector}
The central feature of the CMS detector is a 3.8\unit{T} superconducting solenoid magnet with an inner diameter of 6\unit{m}. Within the magnet volume are the following subdetectors: a silicon pixel and strip tracker, a lead tungstate crystal electromagnetic calorimeter (ECAL), and a brass and scintillator hadron calorimeter. Muons are detected in gas-ionization chambers embedded in the steel flux-return yoke outside the solenoid. In addition, two steel and quartz-fiber hadron forward calorimeters extend the detection coverage to regions close to the beam pipe. A more detailed description of the CMS detector, together with a definition of the coordinate system used and the relevant kinematic variables, can be found in Ref.~\cite{Chatrchyan:2008zzk}.
Events of interest are selected using a two-tiered trigger system~\cite{Khachatryan:2016bia}. The first level, composed of custom hardware processors, uses information from the calorimeters and muon detectors to select events at a rate of around 100\unit{kHz} within a fixed time interval of about 4\mus. The second level, known as the high-level trigger, consists of a farm of processors running a version of the full event reconstruction software optimized for fast processing, and reduces the event rate to around 1\unit{kHz} before data storage.

\section{Simulated data samples}
\label{sec:Data and simulated samples}
Monte Carlo (MC) event generators are used to simulate the SM background processes and the signal. These simulations are used to guide the design of the analysis, to estimate minor backgrounds, and to interpret the results.

Background events are generated at leading order (LO) for the $\PW+\text{jets}$ and $\PZ/\PGg^*+\text{jets}$ processes using the generator~\MGvATNLO 2.2.2 (2.4.2)~\cite{Alwall:2014hca} for simulated events matched with 2016 (2017--18) data, while the next-to-LO (NLO) generator \POWHEG~2.0~\cite{Nason:2004rx, Frixione:2007vw, Alioli:2010xd, Alioli:2011as,Re:2010bp,Chiesa:2020ttl} is used for \ttbar, $\PQt\PW$, and diboson processes, and \sloppy{\MGvATNLO} at NLO for $\ttbar+\PW$, $\ttbar+\PZ/\PGg^*$, $\ttbar\ttbar$, $\PQt\PZ\PQq$, and triboson production. 
Both \MGvATNLO and \POWHEG are interfaced with \PYTHIA 8.226 (8.230)~\cite{Sjostrand:2014zea} for parton showering and hadronization using the tune \textsc{CUETP8M1}~\cite{Khachatryan:2015pea} or \textsc{CUETP8M2T4}~\cite{CMS-PAS-TOP-16-021} (\textsc{CP5}~\cite{Sirunyan:2019dfx}) and the
NNPDF 3.0~\cite{Ball:2014uwa} (3.1~\cite{Ball:2017uwa}) parton distribution functions (PDFs) for simulating all 2016 (2017--18) samples.
In the following, we group these backgrounds where a genuine $\tau$ lepton is present as either ``t production'' or ``Others'', depending on whether a top quark is produced in the SM process or not.

Signal samples are generated at LO using \textsc{\MGvATNLO} interfaced with \textsc{\PYTHIA} for the \LQS and \LQV models of Refs.~\cite{Alvarez:2018gxs} and~\cite{Buttazzo:2017ixm}, according to the implementations of Refs.~\cite{Buchmuller1987442} and~\cite{Dorsner:2018ynv}.
The NNPDF 3.0~\cite{Ball:2014uwa} (3.1~\cite{Ball:2017uwa})
parton distribution function (PDF) set is utilized with the tune \textsc{CUETP8M1}~\cite{Khachatryan:2015pea} (\textsc{CP2}~\cite{Sirunyan:2019dfx}) for the signal events used with the 2016 (2017--18) data.
The \LQ mass range studied is between 0.5 and 2.3\TeV, with samples produced in steps of 0.3\TeV. 
We consider \LQS (\LQV) decaying as $\LQ\to\PQt\tau$ ($\PQt\nu$) or $\LQ\to\PQb\nu$ ($\PQb\tau$).
Samples of pair-produced leptoquarks are generated considering both gluon-initiated and quark-initiated mechanisms.
We consider equal values of $\lambda$ for \LQs coupled to ($\PQt\PGt$, $\PQb\PGn$) and ($\PQt\PGn$, $\PQb\PGt$).
Samples of singly produced \LQ are generated with $\lambda$ values 0.1, 0.5, 1, 1.5, 2, and 2.5.
In the MC simulation, the kinematic distributions of singly produced leptoquarks are independent of $\lambda$ below $\lambda = 0.5$ (1) in the case of \LQS (\LQV), and in both cases are independent of $\emph{k}$.
The dependence on $\lambda$ above these values is ascribed 
to the contributions of virtual \LQ states in the quark-gluon fusion amplitude (Fig.~\ref{fig:LQtopFeynmanDiagram} right) that become more and more relevant compared to the resonant \LQ production for increasing values of \LQ mass and $\lambda$, and are manifest as off-shell events that tend to populate the low-mass tail. 

Additional $\Pp\Pp$ interactions within the same or nearby bunch crossings
(pileup) are taken into account by superimposing simulated minimum bias interactions onto the hard scattering process, with a number distribution matching that observed in data. Simulated events are propagated through the full GEANT4 based simulation~\cite{Geant} of the CMS detector.

\section{Particle reconstruction and identification}
\label{sec:Object reconstruction}
A particle-flow (PF) algorithm~\cite{Sirunyan:2017ulk} is used to identify and reconstruct individual particles in the event (electrons, muons, photons, neutral and charged hadrons) through a combination of the information from the entire detector. These PF objects are used to reconstruct higher-level objects such as hadronically decaying $\tau$ leptons (\tauh), jets, and missing transverse momentum ($\ptvecmiss$), taken as the negative vector sum of the transverse momenta (\ptvec) 
of all reconstructed particles in an event. The magnitudes of \ptvec and \ptvecmiss are referred to as \pt and \ptmiss\unskip, respectively.

Jet candidates are reconstructed from PF candidates using the anti-\kt clustering algorithm~\cite{Cacciari:2008gp} with a distance parameter of 0.8 (``AK8 jet'') or 0.4 (``AK4 jet''), and are selected requiring $\pt > 30\GeV$ and $\abs{\eta} < 2.4$.
The jet energy scale (JES) is calibrated through correction factors dependent on the \pt, pseudorapidity ($\eta$), energy density, and the area of the jet. The jet energy resolution (JER) for the simulated jets is corrected to reproduce the resolution observed in data~\cite{Khachatryan:2016kdb}.

The AK8 jet candidates are required to have $ \pt > 180\GeV$, $\abs{\eta} < 2.4$, and to be separated by $\Delta R > 0.8$ from an identified \tauh, where $\Delta R \equiv \sqrt{\smash[b]{(\Delta\eta)^2+(\Delta\phi)^2}}$ and $\phi$ is the azimuthal angle. 
They are selected if they are identified as originating from a \PW boson decaying to $\PQq\PAQq^{\prime}$ (denoted as ``\PW jets'') by using a pruning algorithm~\cite{Ellis:2009su} or from a top quark decaying fully hadronically (``\PQt jets'').
The mass of the pruned AK8 \PW jet is required to be within the range 65--105\GeV to select candidates consistent with \PW bosons and to reject quark and gluon jets. The discrimination between \PW jets and quark and gluon jets is further improved by requiring the ratio $\tau_{21}$ to be less than 0.35 or 0.45, depending on the year of data taking, where  $\tau_{21} \equiv \tau_2 / \tau_1$ and the $N$-subjettiness $\tau_n$ has the property that it attains a smaller value the more nearly the jet resembles a collection of $n$ subjets~\cite{Khachatryan:2014vla}.
In a similar way, an AK8 jet may be identified as arising from the fully hadronic decay of a top quark. These \PQt jets are required to have $ \pt > 400\GeV$, mass of the jet reconstructed through the modified mass drop tagger algorithm~\cite{Dasgupta:2013ihk, Larkoski:2014wba} between 105 and 220\GeV, and $\tau_{32} \equiv \tau_3 / \tau_2$ less than 0.81.

The \tauh candidates are reconstructed with the hadron-plus-strips algorithm~\cite{Sirunyan:2018pgf}, which is
seeded with AK4 jets. 
This algorithm reconstructs \tauh candidates in the one-prong, one-prong plus {\PGpz}(s), and three-prong decay modes. 
A discriminator based on a multivariate analysis, including isolation~\cite{Sirunyan:2018pgf} as well as lifetime information, is used to reduce the frequency of jets
being misidentified as \tauh candidates. The typical working point used in this analysis has an efficiency of ${\approx} 60$\% for a genuine \tauh, with a misidentification rate for quark and gluon jets of ${\approx} 0.1$\%~\cite{Sirunyan:2018pgf}. Electrons and muons misidentified as \tauh candidates are suppressed using criteria based on the consistency among the measurements in the tracker, the calorimeters, and the muon detectors. The \tauh candidates are required to have a minimum \pt of 20 GeV and $\abs{\eta}<2.3$.

Jets arising from a bottom quark (``\PQb jets'') are identified among AK4 jets using the combined secondary vertex algorithm~\cite{Sirunyan:2017ezt}. 
We choose a ``loose'' working point that has an efficiency of 85\% for genuine \PQb jets and a rejection of 90\% of light-flavor jets.
The \PQb jets are considered regardless of whether they are contained in top quark candidates.  

Further requirements are imposed on the AK4 jets used in the construction of top and bottom quark candidates. 
These are required to have $\pt > 30\GeV$ and $\abs{\eta} < 2.4$, and to be separated by $\Delta R > 0.4$ (0.8) from an identified \tauh (\PW jet).

A hadronically decaying top quark candidate is reconstructed considering three cases: an AK8 jet identified as a \PQt jet, a pair comprising an AK4 jet and a \PW jet
and having combined mass closest to the top quark mass among such pairs, and the triplet of AK4 jets having a mass closest to the top quark mass. 
The \PQb tagging information is not used in any of these three reconstruction processes.
These cases correspond to the three possible topologies of hadronic top quark decay and are referred to as ``fully merged'', ``partially merged'', and ``resolved'', respectively.  The reconstruction considers these cases in the order just described, removing the objects contained in a candidate from further consideration to ensure that the categories are exclusive.  
The efficiency for identifying \PW, \PQb, and \PQt jets in simulation is corrected to match the results found in data~\cite{Sirunyan:2020lcu,Sirunyan:2017ezt}.

To select events from processes with fully hadronic states, a veto on electrons and muons is applied.
Electron candidates are reconstructed by combining the information from the ECAL and the silicon tracker, and are identified if they satisfy quality requirements and isolation as specified in~\cite{Khachatryan:2015hwa}; they are selected if they have $\pt > 20\GeV$ and $\abs{\eta} < 2.5$.
Muon candidates are reconstructed by combining the information from 
the muon system and the silicon tracker, and are identified if they pass additional identification criteria and isolation as specified in~\cite{Sirunyan:2018fpa}; they are selected if they have $ \pt > 20\GeV$ and $\abs{\eta} < 2.4$.

\section{Event selection}
\label{sec:Event selection}
Selected events must satisfy a trigger that requires both \ptmiss and \mht greater than 120\GeV, \mht being the magnitude of the negative summed \ptvec of all the AK4
jets reconstructed with the PF algorithm. 

Offline, we consider events in which both $\ptmiss$ and $\mht\ge 200\GeV$, and $\HT\ge 300\GeV$, where \HT is the scalar sum of the \pt of all AK4 jets. Events entering this region are further required to contain exactly one top quark candidate, one \tauh candidate, no electrons or muons, and at least one \PQb jet. 
Finally, the transverse mass~$\mT(\tauh,\ptmiss) \equiv \sqrt{\smash[b]{2 \pt(\tauh) \ptmiss [1-\cos \Delta\phi(\ptvec(\tauh),\ptvecmiss)]}}$ has to exceed 300 \GeV, where $\ptvec(\tauh)$ is the transverse momentum vector of the $\tauh$ candidate. 

From simulation we find that the total selection efficiency, accounting for both the \LQ decay branching fraction and the event selection, varies between about 2 and 9\% for an \LQ mass in the range 0.5--2.3\TeV for pair-produced \LQs. 
For singly produced \LQs, taking $\lambda = 1.5$, the signal efficiency is about 0.7\% for an \LQS with a mass of 1.1\TeV; the corresponding number for an \LQV is 2.4 (3.1)\% for $k = 0$ (1) for a mass of 1.4\TeV. 
The efficiency decreases for higher $\lambda$ and \LQ mass values. This is because of the increased impact of the virtual leptoquarks leading to the nonresonant process in which the events tend to populate the low-mass tail, as described in Section~\ref{sec:Data and simulated samples}.
The efficiency values for all the different leptoquark hypotheses and parameters investigated in this search can be found in the HEPData database~\cite{hepdata}.
The search is less sensitive to single \LQ production than to pair production because of the smaller signal efficiency for higher $\lambda$ and \LQ mass values and the similarity of the kinematic properties to those of the expected SM background. These effects outweigh the higher relative \LQS (\LQV $k = 0$, \LQV $k = 1$) cross section for mass values of 0.5 and 0.7 \TeV (0.6 and 1.2 \TeV, 1.2 and 2 \TeV) at values of $\lambda$ of 2 and 1.5.

\begin{figure*}[!th]
\centering
\includegraphics[width=0.49\textwidth]{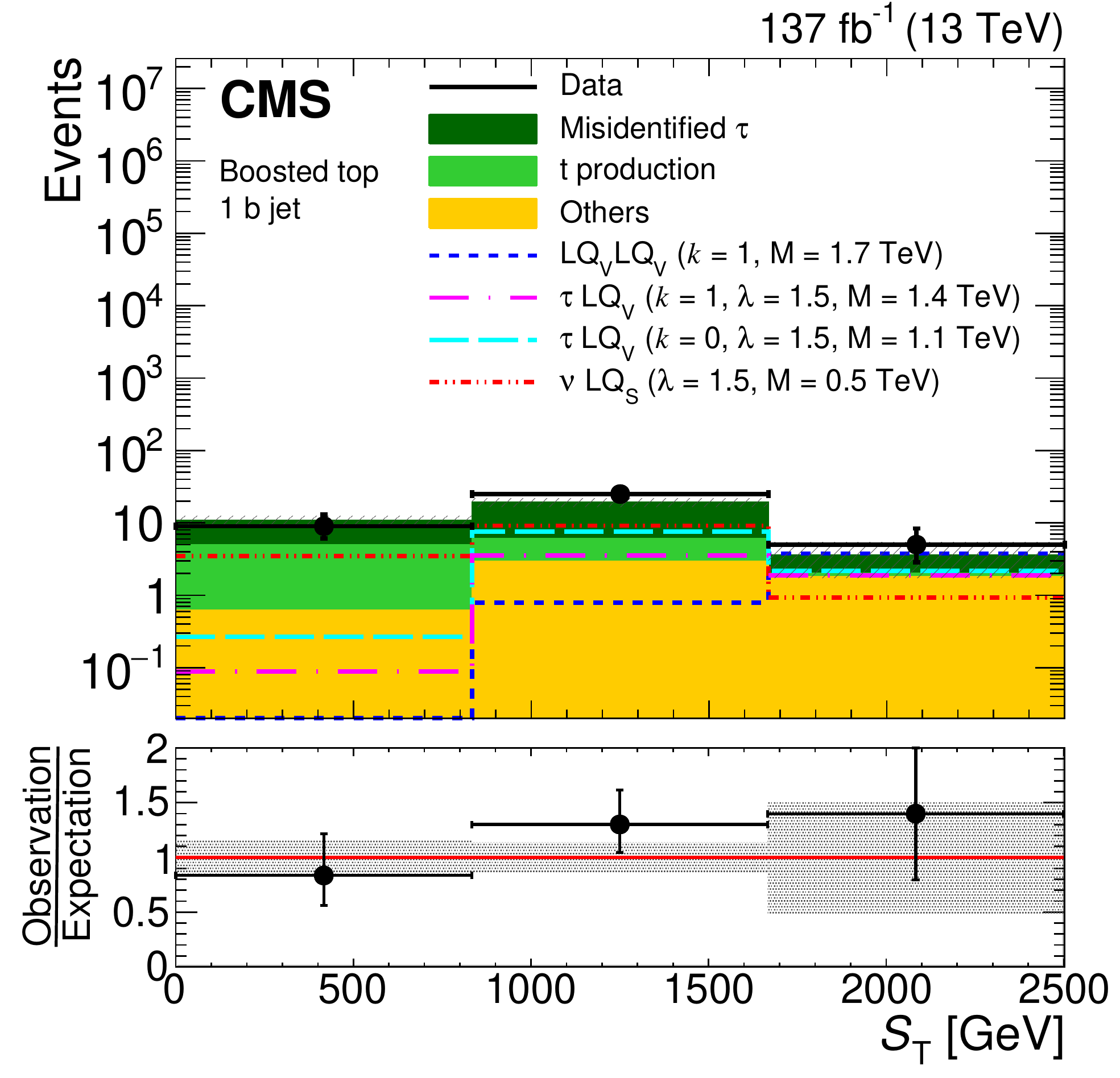}
\includegraphics[width=0.49\textwidth]{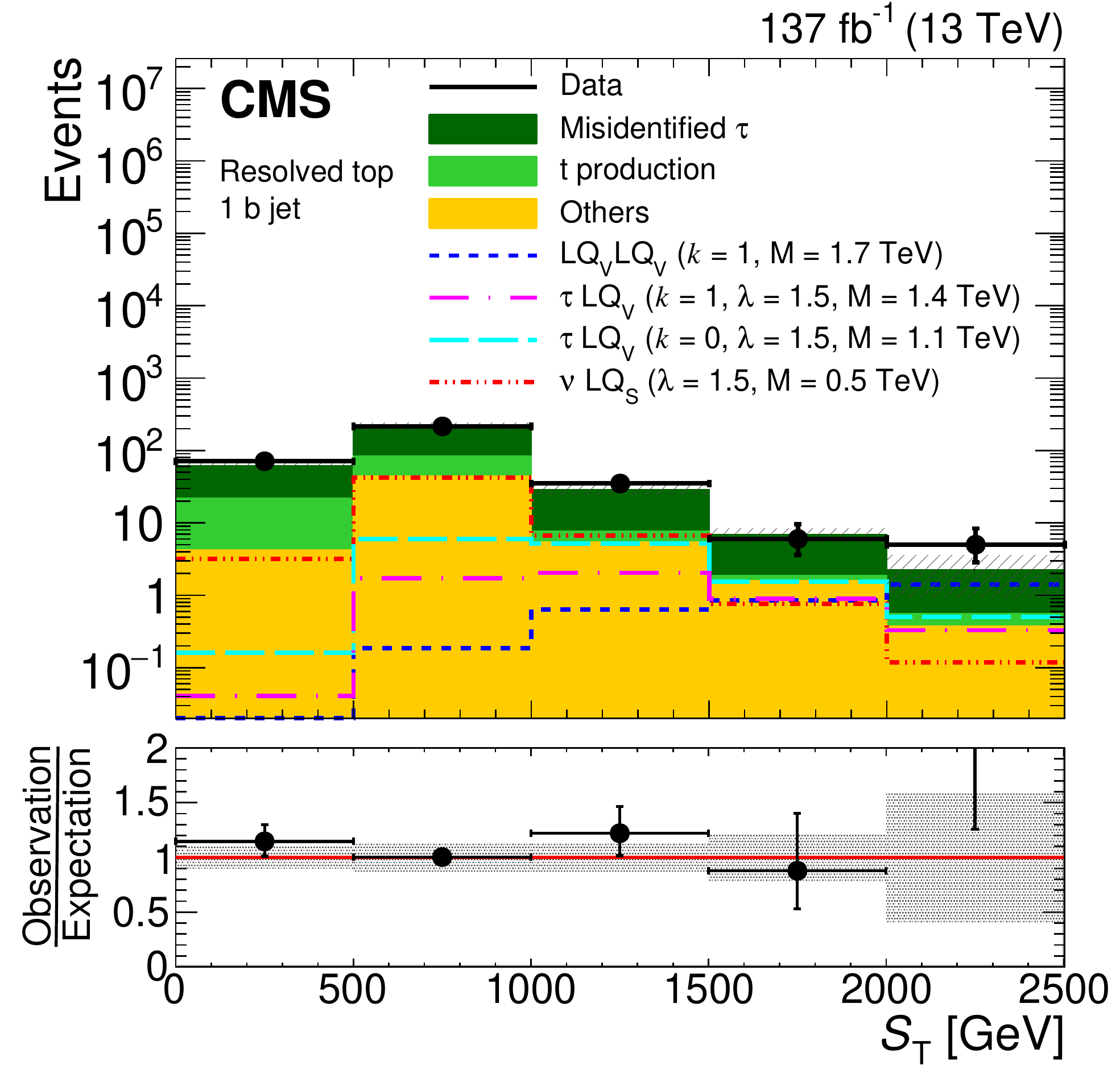}\\
\includegraphics[width=0.49\textwidth]{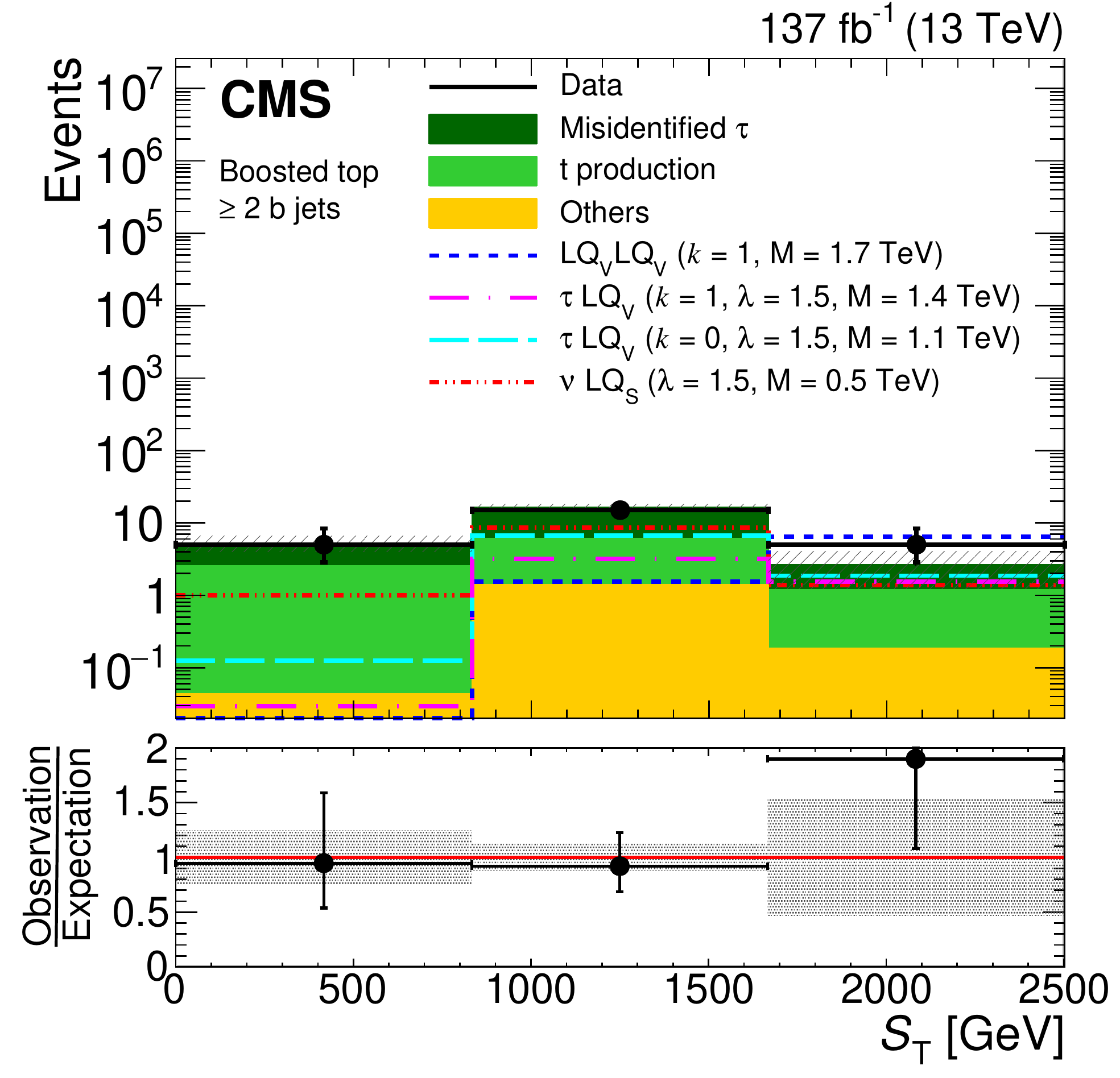}
\includegraphics[width=0.49\textwidth]{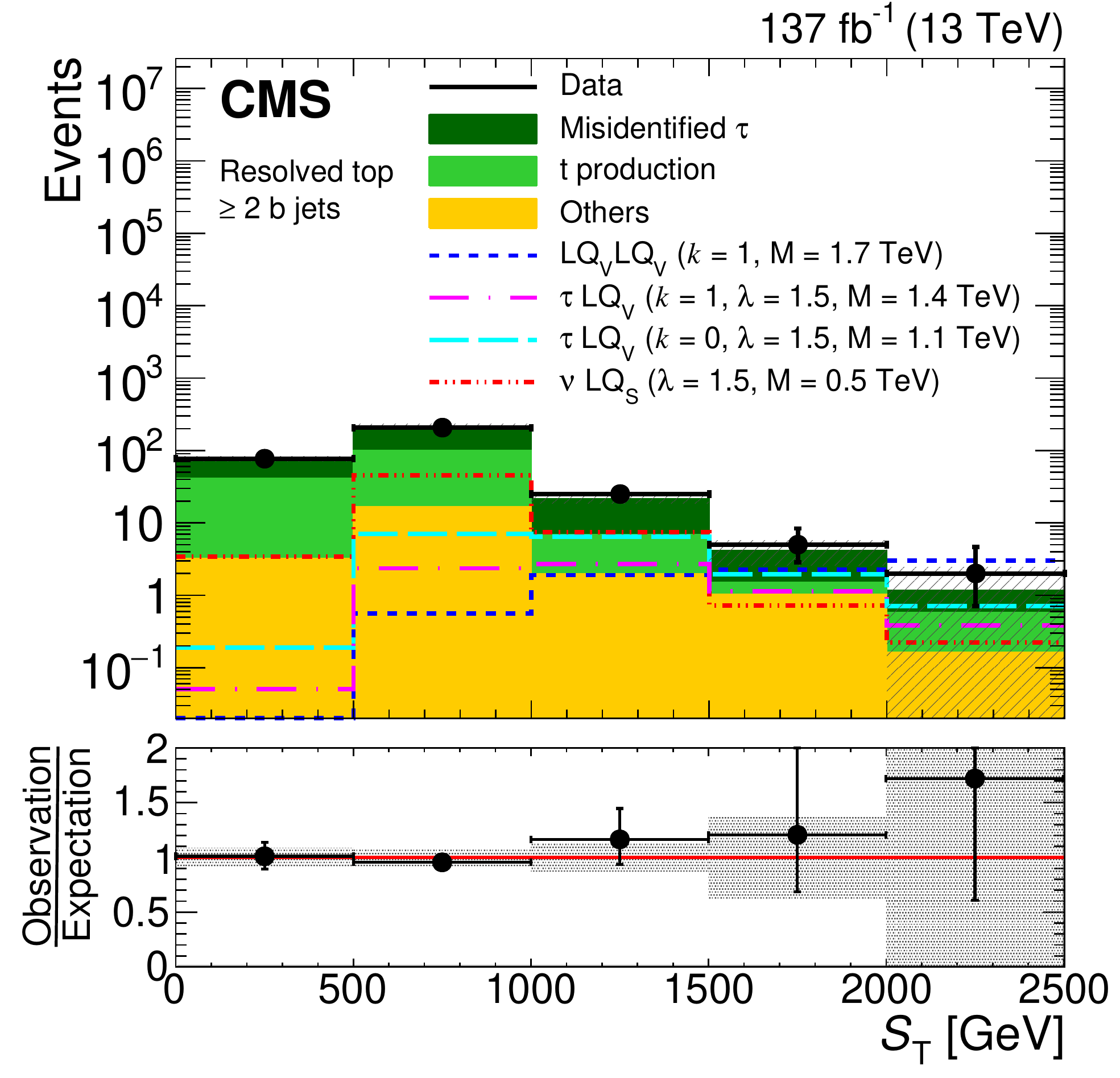}\\
\caption{Distribution of the variable \ST for events passing the signal selection for the SM background estimation (stacked filled histograms), data (black points), and different hypotheses of \LQ signals (lines). Upper left: boosted top quark candidate (hadronically decaying top quark reconstructed in the fully or partially merged topology) and exactly one \PQb jet; lower left: boosted top quark candidate and at least two \PQb jets; upper right:  resolved top quark candidate (hadronically decaying top quark reconstructed in the resolved topology) and exactly one \PQb jet; lower-right: resolved top quark candidate and at least two \PQb jets. The cross-hatched band in the upper panels represents the total uncertainty (statistical+systematic). The lower panel of each distribution shows the ratio, and its uncertainty, between the observation and the SM expectation. 
}\label{fig:sigSelection}
\end{figure*}

The events that pass the above selection are categorized according to the number of \PQb jets ($\nbjets = 1$ or $\ge2$)
and to whether the top quark candidate is selected through the fully or partially merged topology (``boosted''), or the resolved topology (``resolved''). 
For each of these four categories of events
a distribution-based analysis is performed, searching for evidence of a signal by considering the distribution of \ST, which is the scalar sum of the \pt of the top quark candidate, the selected \tauh, and the $\ptmiss$. Figure~\ref{fig:sigSelection} shows the \ST distributions for the events passing the signal selection in the four categories of the analysis, while Table~\ref{tab:sigSelection} gives the yields from the background estimation and the expected signal.

\begin{table*}[th]
\centering
  \topcaption{
Yields from the SM background estimation, data, and expected signal, for the selected events, with total (statistical+systematic) uncertainties.
}\label{tab:sigSelection}
\setlength\tabcolsep{0.5pt}
\begin{tabular}{lrllrllrllrll}
\hline
\multirow{3}{*}{Category} & \multicolumn{6}{c}{Boosted}  & \multicolumn{6}{c}{Resolved} \\
                     & $\nbjets$ &$=$& $1$ \hspace{0.75cm} &$\nbjets$ &$\ge$& $2$ \hspace{0.75cm} &$\nbjets$ &$=$& $1$ \hspace{0.75cm} &$\nbjets$ &$\ge$& $2$ \\ \hline
Misidentified \PGt   & 20.5  & $\pm$ & 2.1  & 14.4  & $\pm$ & 1.8  & 199   & $\pm$ & 13   & 170   & $\pm$ & 12  \\
\PQt production      & 7.8   & $\pm$ & 2.1  & 8.2   & $\pm$ & 1.9  & 59    & $\pm$ & 5    & 127   & $\pm$ & 10 \\
Others               & 5.3   & $\pm$ & 2.0  & 1.6   & $\pm$ & 0.8  & 56    & $\pm$ & 25   & 23    & $\pm$ & 11  \\[\cmsTabSkip]
Total background     & 33.5  & $\pm$ & 3.6  & 24.2  & $\pm$ & 2.7  & 314   & $\pm$ & 29   & 320   & $\pm$ & 19  \\[\cmsTabSkip]
Data                 & 39    &       &      & 25    &       &      & 332   &       &      & 316   &       &      \\[\cmsTabSkip]
\LQV\ALQV ($k = 1$, $m_{\mathrm{LQ}} = 1.7\TeV$)   
                     & 4.6   & $\pm$ & 0.7  & 8.0   & $\pm$ & 1.2  & 3.1   & $\pm$ & 0.3  & 7.7   & $\pm$ & 0.7 \\
$\tau$\LQV ($k = 1$, $\lambda = 1.5$, $m_{\mathrm{LQ}} = 1.4\TeV$) \hspace{0.2cm} 
                     & 5.5   & $\pm$ & 0.4  & 4.8   & $\pm$ & 0.4  & 5.0   & $\pm$ & 0.2  & 6.6   & $\pm$ & 0.3\\
$\tau$\LQV ($k = 0$, $\lambda = 1.5$, $m_{\mathrm{LQ}} = 1.1\TeV$)                
                     & 10.1  & $\pm$ & 0.7  & 8.6   & $\pm$ & 0.7  & 13.4  & $\pm$ & 0.6  & 16.4  & $\pm$ & 0.8\\
$\nu$\LQS ($\lambda = 1.5$, $m_{\mathrm{LQ}} = 0.5\TeV$)                
                     & 13.5  & $\pm$ & 0.8  & 12.0  & $\pm$ & 0.8  & 52.7  & $\pm$ & 2.7  & 57.5  & $\pm$ & 2.9
\\\hline
\end{tabular}
\end{table*}

\section{Background estimation}
\label{sec:Background estimation}
Several SM processes contribute as backgrounds in the signal region. We treat separately the two cases in which a genuine $\tau$ lepton is present or not in the event. 

The irreducible background with a real $\tau$ lepton that decays hadronically is estimated from simulated samples, 
and normalized to data in a control region where we expect negligible contribution from the signal to account for residual differences between data and simulation. 
Processes with at least one top quark (e.g. \ttbar or $\ttbar+\PW$) account for most of this irreducible background, and a control region is defined by applying the requirements used for the signal region, except with $\mT(\tauh,\ptmiss) < 80\GeV$ and $\nbjets\ge2$.

The dominant source of contamination is the reducible background, which comprises all of the processes 
(mainly events composed uniquely of jets produced through the strong interaction, $\PW+\text{jets}$, and \ttbar) 
that pass the signal region selection and in which a jet is misidentified as a \tauh candidate. 
We estimate this background entirely from data by applying misidentification weights $w$ to the yields of events selected with the same requirements as the signal region, except that the \tauh must pass a looser identification requirement and fail the nominal one.  We refer to this sample as the application region.
An estimate from simulation of the number of events entering the application region while having a genuine \tauh is subtracted from the application region yields.
The weight $w$ of each event depends on the probability $f$ that a misidentified \tauh candidate passing the relaxed criteria also passes the nominal criteria, and is given by $f/(1-f)$.
The probability $f$ is parameterized as a function of the \pt and $\abs{\eta}$ of the jet associated with the selected \tauh candidate, within $\Delta$R$(\text{jet},\tauh)<0.4$.  It is measured in a large data sample with a high fraction of jets misidentified as \tauh.
To select this sample the signal region requirements are modified by removing the thresholds on \ptmiss and \mht and requiring instead the presence of a muon with \pt greater than 60\GeV.
The requirement $\nbjets \ge 1$ is replaced by $\nbjets = 0$, to suppress \ttbar events with genuine \tauh, and the requirement on $\mT(\tauh,\ptmiss)$ is replaced by $\mT(\tauh,\PGm)>120\GeV$, to suppress Drell--Yan events. In the resultant sample, more than 90\% of the events have jets misidentified as \tauh, with W+jets contributing 60\% and the rest consisting of a mixture of top, diboson, and multijet events.
This estimation method has been validated in a region that passes the signal region selection, except for the modified requirement $120 < \mT(\tauh,\ptmiss) < 300\GeV$.
This region is verified to have a composition of background processes similar to that of the signal region but is dominated by events with a misidentified \tauh candidate, as determined from MC simulation. We find good agreement between the data and the estimated background in this region, as well as in a larger one with the $\nbjets$ requirement released. The observed difference does not exceed 12\%, and this value is therefore assigned as the systematic uncertainty in the background estimated using this method.

\section{Systematic uncertainties}\label{sec:systematics}
Systematic uncertainties from various sources are propagated to both the shape and normalization of the distributions in the discriminating variable \ST.
The systematic uncertainties affect both the signal and the background, particularly the minor backgrounds (t production or ``Others'') that are derived relying on the MC simulation, while the main background ($\PGt$ misidentification) is estimated from data.

The shape uncertainties vary according to the background process, \ST bin, and year of data taking. Thus in the following, we quote a range of values, reflecting the minimum and maximum uncertainties observed under the various conditions. The effect of the uncertainty on the simulation of pileup is estimated by varying the inelastic cross section~\cite{Sirunyan:2018nqx} used in the simulation by 5\%. This results in an uncertainty associated with the background of between 1 and 
6\%, and of 1\% associated with the signal.
The uncertainty in the acceptance associated with the PDFs is evaluated in accordance with the PDF4LHC recommendations~\cite{PDF4LHC}, using the PDF4LHC15 Hessian PDF set with 100 eigenvectors, and is found to be less than 5\% for the signal. The uncertainty related to the trigger is between 1 and 
2\%, for both the background and the signal. The jet four-momenta are varied within the JES and the JER uncertainties~\cite{Khachatryan:2016kdb}, resulting in an effect that ranges between 1 and 35\% for the background and up to 2.5\% for the signal. The above uncertainties are correlated across the years, while those discussed below are treated as uncorrelated, as they are dominated by statistical uncertainties. Corrections related to the \PQb tagging are varied by the uncertainties that are measured with control samples in data and simulation~\cite{Sirunyan:2017ezt}, giving a systematic uncertainty in the yields in the range 3--10\% for the background and 8--10\% (13--23\%) for single (pair) \LQ production. Analogously, we take into account the uncertainty in the \tauh energy scale and identification~\cite{Sirunyan:2018pgf}, which amounts to 1--5\% (less than 1\%) and 5--13 (13--20)\% for the background (signal). The \PW and \PQt jet tagging uncertainty amounts to 2--11 (1--4)\% and 3--15 (7--14)\% for the background (signal). 
For all of the background processes, the statistical uncertainty in the samples used is included in the systematic uncertainty. 

The sources of systematic uncertainty that affect only the normalization are the uncertainties in the cross sections of the backgrounds estimated from simulation (5\% for top quark production and 30\% for the remaining backgrounds), the uncertainty in the misidentified \tauh contribution, whose value of 12\% is assigned from the consistency test discussed in Section~\ref{sec:Background estimation}, and the uncertainty in the integrated luminosity. The integrated luminosities of the 2016--18 data-taking periods are individually known with uncertainties in the 2.3--2.5\% range~\cite{CMS-PAS-LUM-17-001,CMS-PAS-LUM-17-004,CMS-PAS-LUM-18-002}, while the total Run~2 (2016--18) integrated luminosity has an uncertainty of 1.8\%, the improvement in precision reflecting the (uncorrelated) time evolution of some systematic effects.

\section{Results}\label{sec:results}
Figure~\ref{fig:sigSelection} and Table~\ref{tab:sigSelection} show that the data are in agreement with the background expectations from the SM in all of the event categories investigated.
We proceed with setting upper limits at 95\% confidence level (\CL) on the cross sections for the production of leptoquarks in pairs, $\sigma(\Pp\Pp \to \LQ\ALQ)$, and singly, $\sigma(\Pp\Pp \to \ell\LQ)$, for \LQS ($\ell = \nu)$, and \LQV ($\ell = \tau$). We use the \CLs criterion~\cite{Junk,Read:2002hq} with binned templates of both background and signal as given by the distributions of Fig.~\ref{fig:sigSelection}. For each category and each bin of \ST, the observed number of events is fitted by a Poisson distribution, whose mean is the sum of the total SM expectation, determined as described in Section~\ref{sec:Background estimation}, and a potential signal contribution determined from simulation. The systematic uncertainties described in Section~\ref{sec:systematics} are considered as nuisance parameters, with a lognormal distribution for the normalization parameters and a Gaussian distribution for systematic uncertainties affecting the shape.

\begin{figure*}[!tbp]
\centering
\includegraphics[width=0.49\textwidth]{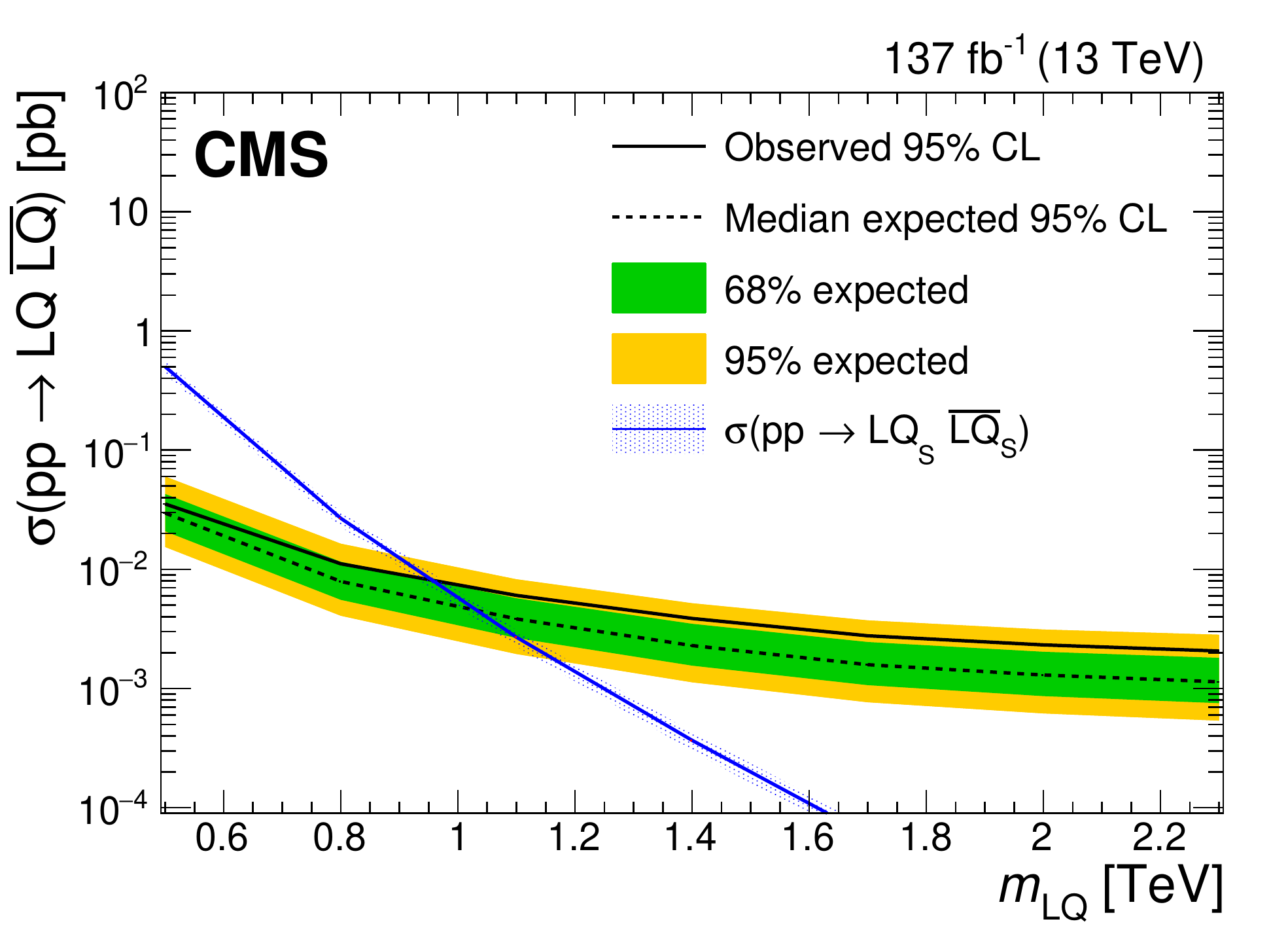}\\
\includegraphics[width=0.49\textwidth]{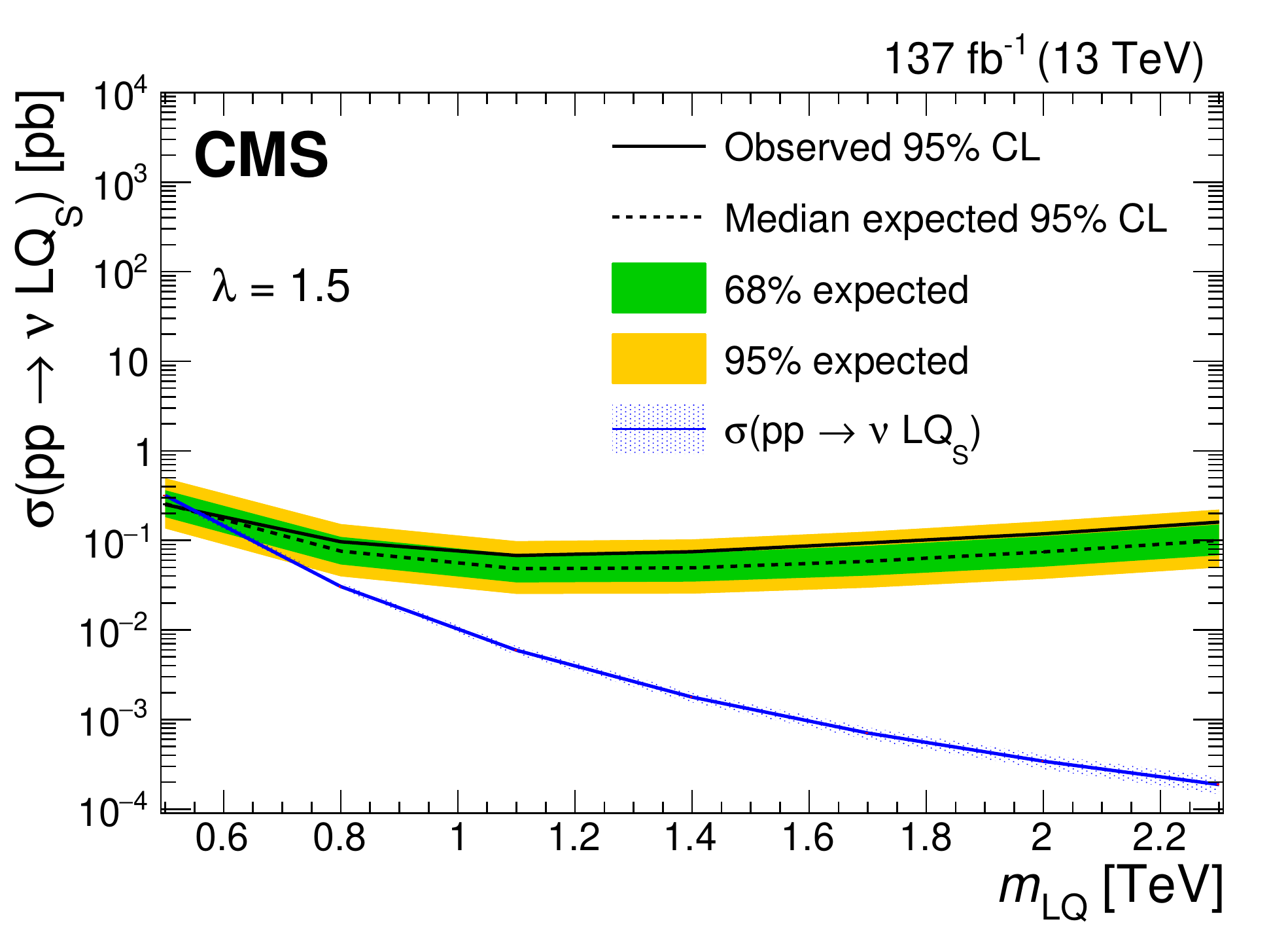}
\includegraphics[width=0.49\textwidth]{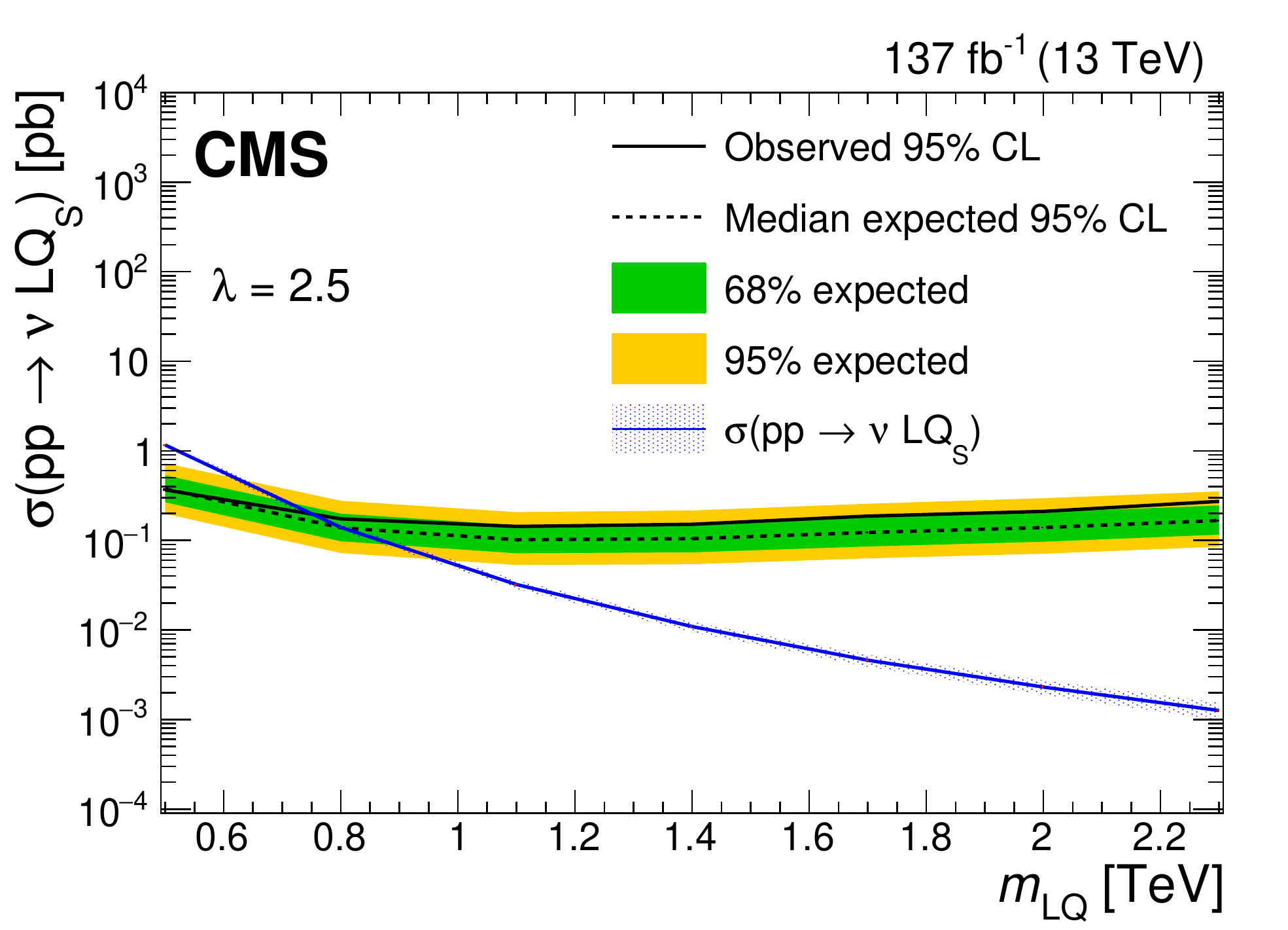}\\
\includegraphics[width=0.49\textwidth]{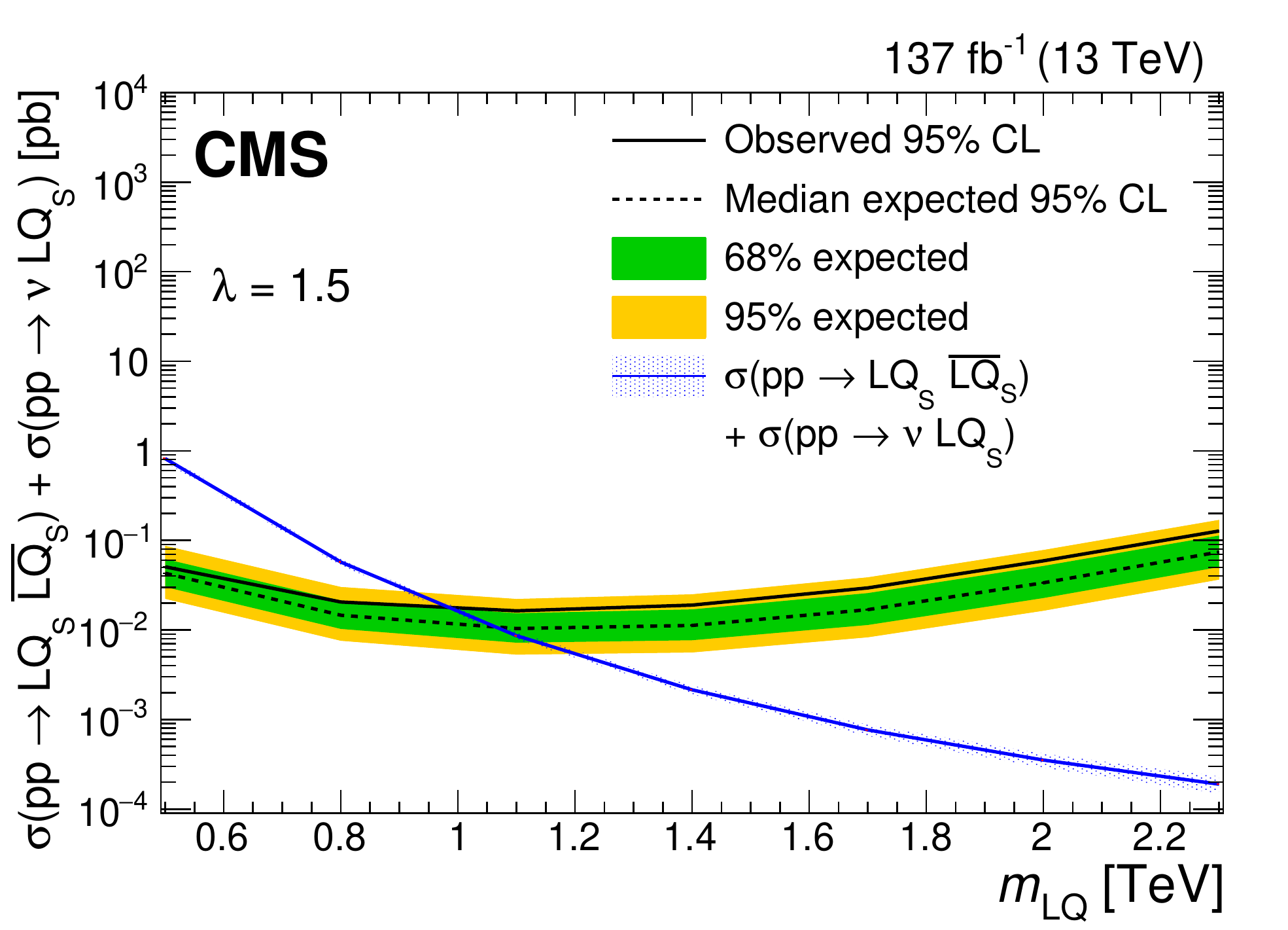}
\includegraphics[width=0.49\textwidth]{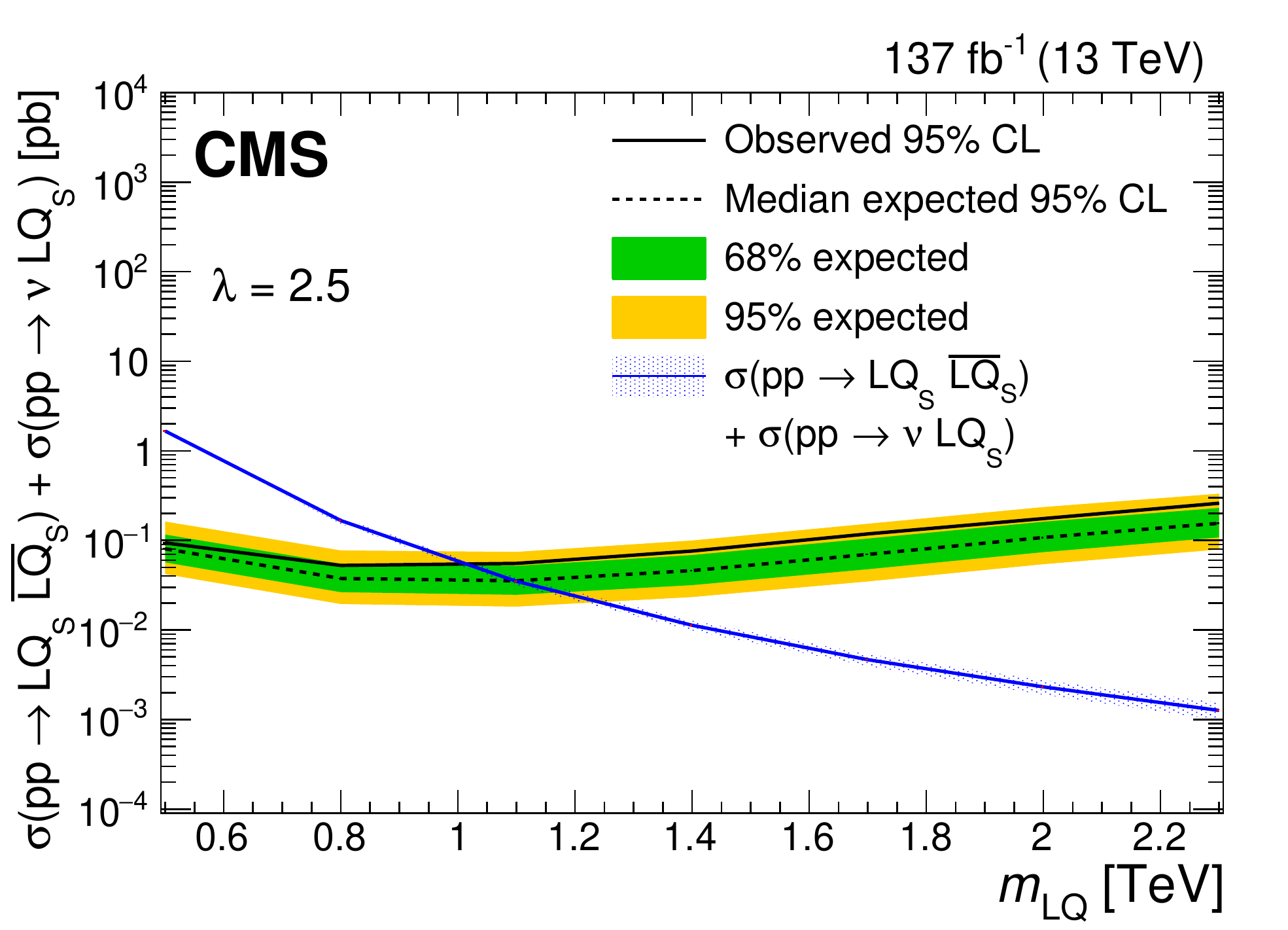}\\
\caption{The observed and expected (solid and dotted black lines) 95\% \CL upper limits on $\sigma(\Pp\Pp\to\LQS\ALQS)$ (upper), $\sigma(\Pp\Pp\to\nu\LQS)$ with $\lambda = 1.5$ and $2.5$ (middle left and right), and $\sigma(\Pp\Pp\to\LQS\ALQS)+\sigma(\Pp\Pp\to\nu\LQS)$ with $\lambda = 1.5$ and $2.5$ (lower left and right), as a function of the mass of the \LQS. The limits apply under the assumption of equal couplings for the \LQ decay to each of the two allowed lepton flavor pairings.
The bands represent the one- and two-standard deviation variations of the expected limit.
The solid blue curve indicates the theoretical predictions at LO, except for pair-produced \LQS, for which NLO values are used based on NLO quantum chromodynamics corrections~\cite{Kramer:2004df} and the model implementation in Ref.~\cite{Dorsner:2018ynv}.}
\label{fig:LimLQs}
\end{figure*}

\begin{figure*}[!tbp]
\centering
\includegraphics[width=0.49\textwidth]{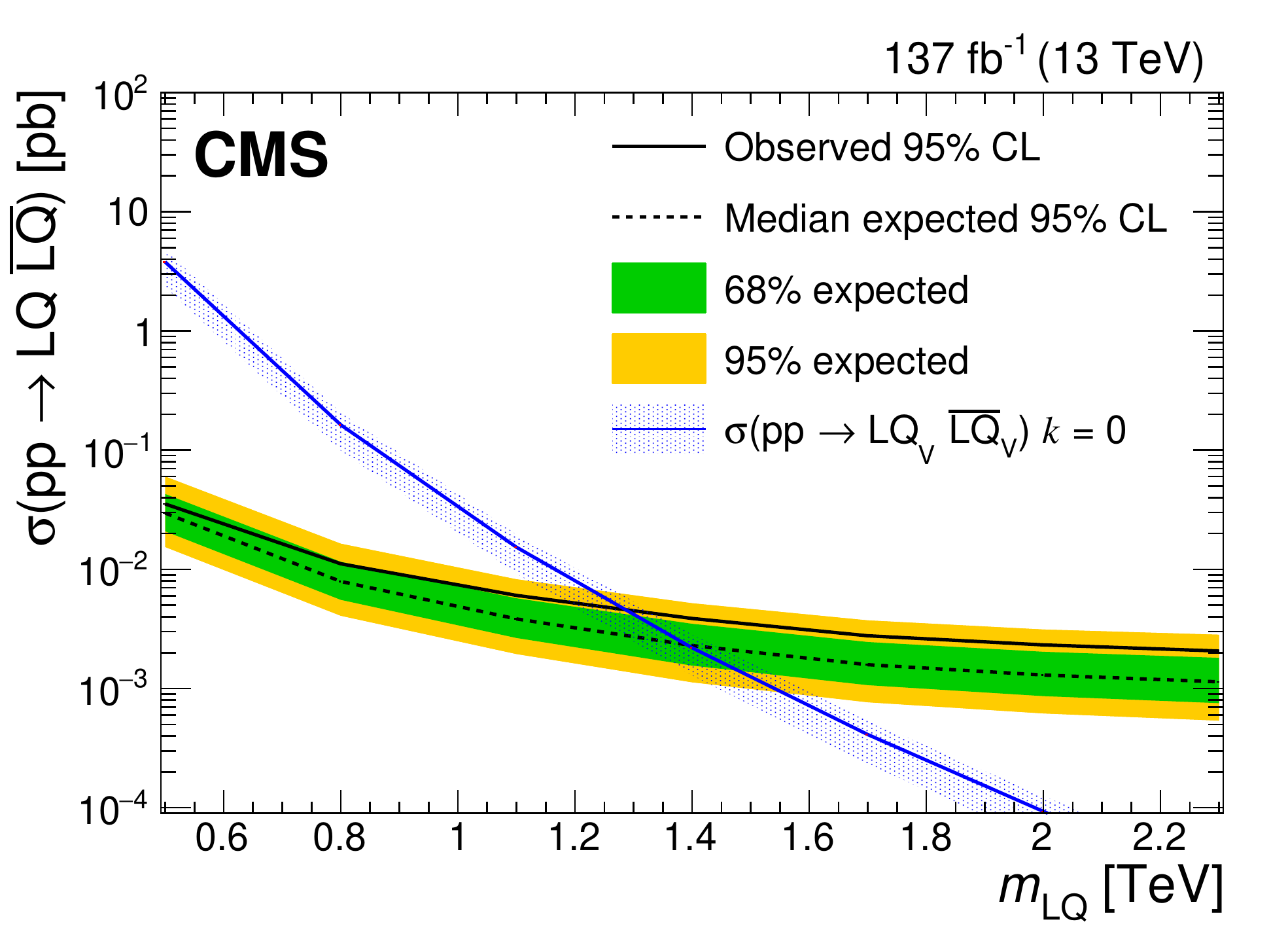}\\
\includegraphics[width=0.49\textwidth]{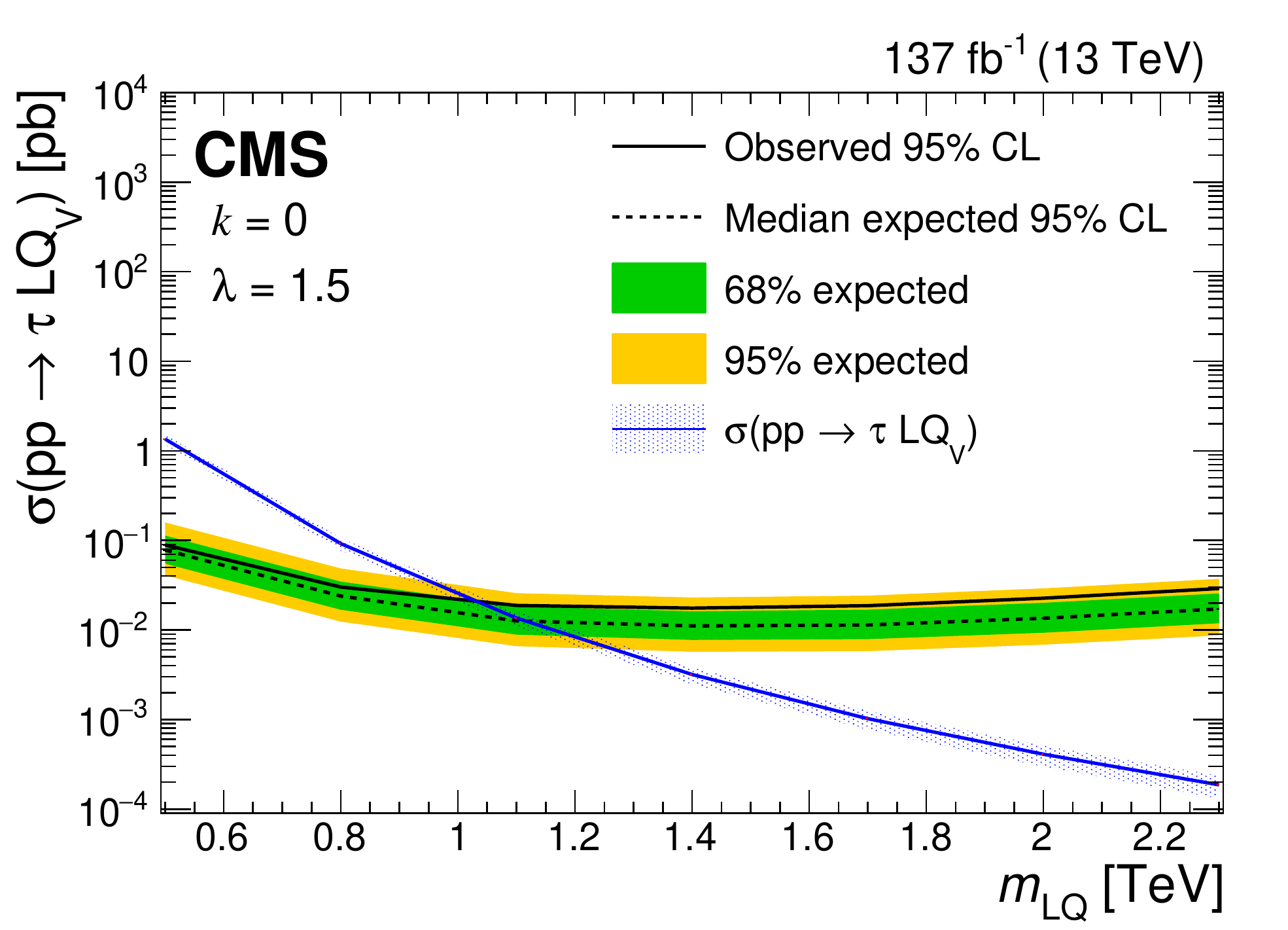}
\includegraphics[width=0.49\textwidth]{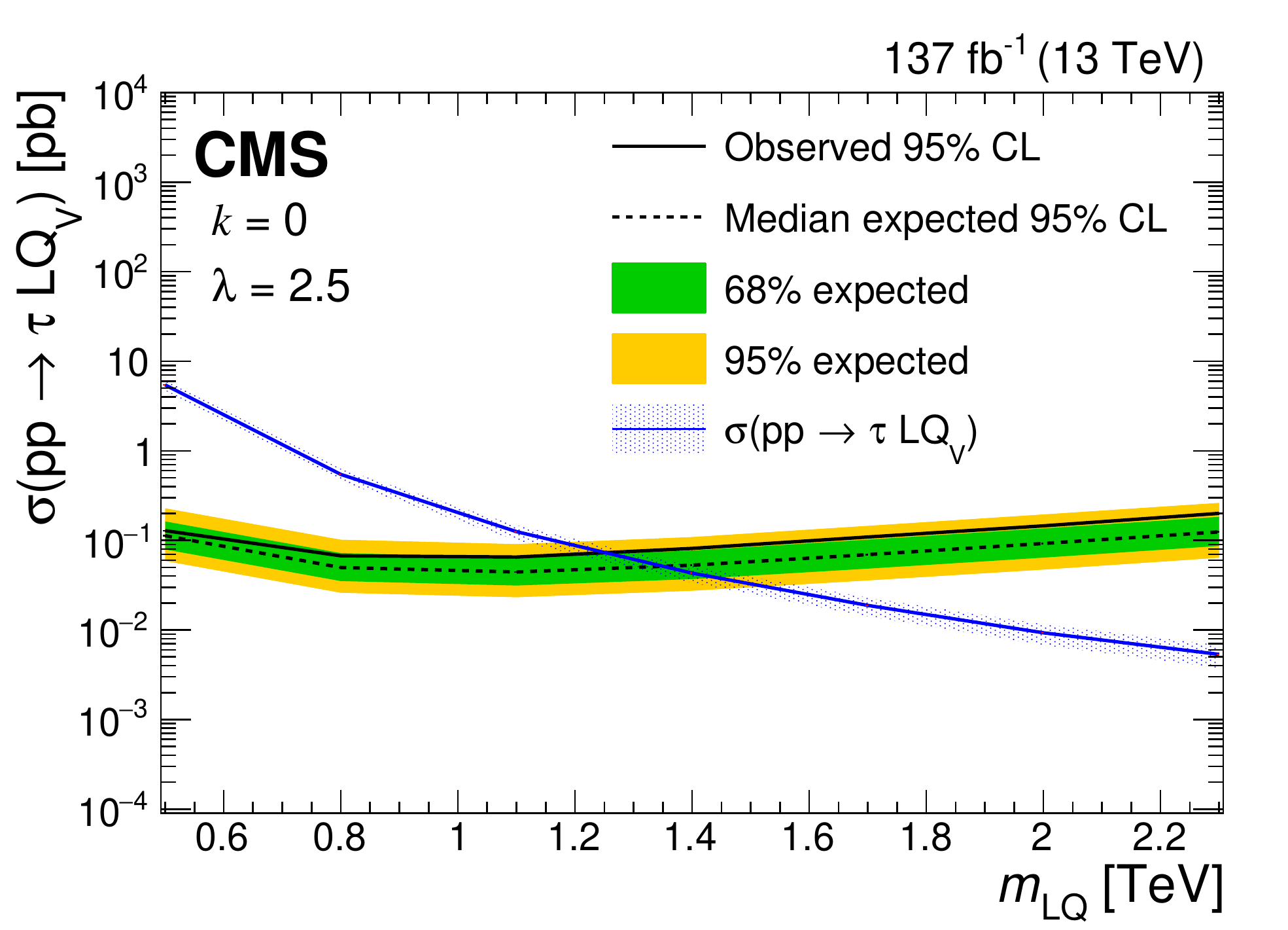}\\
\includegraphics[width=0.49\textwidth]{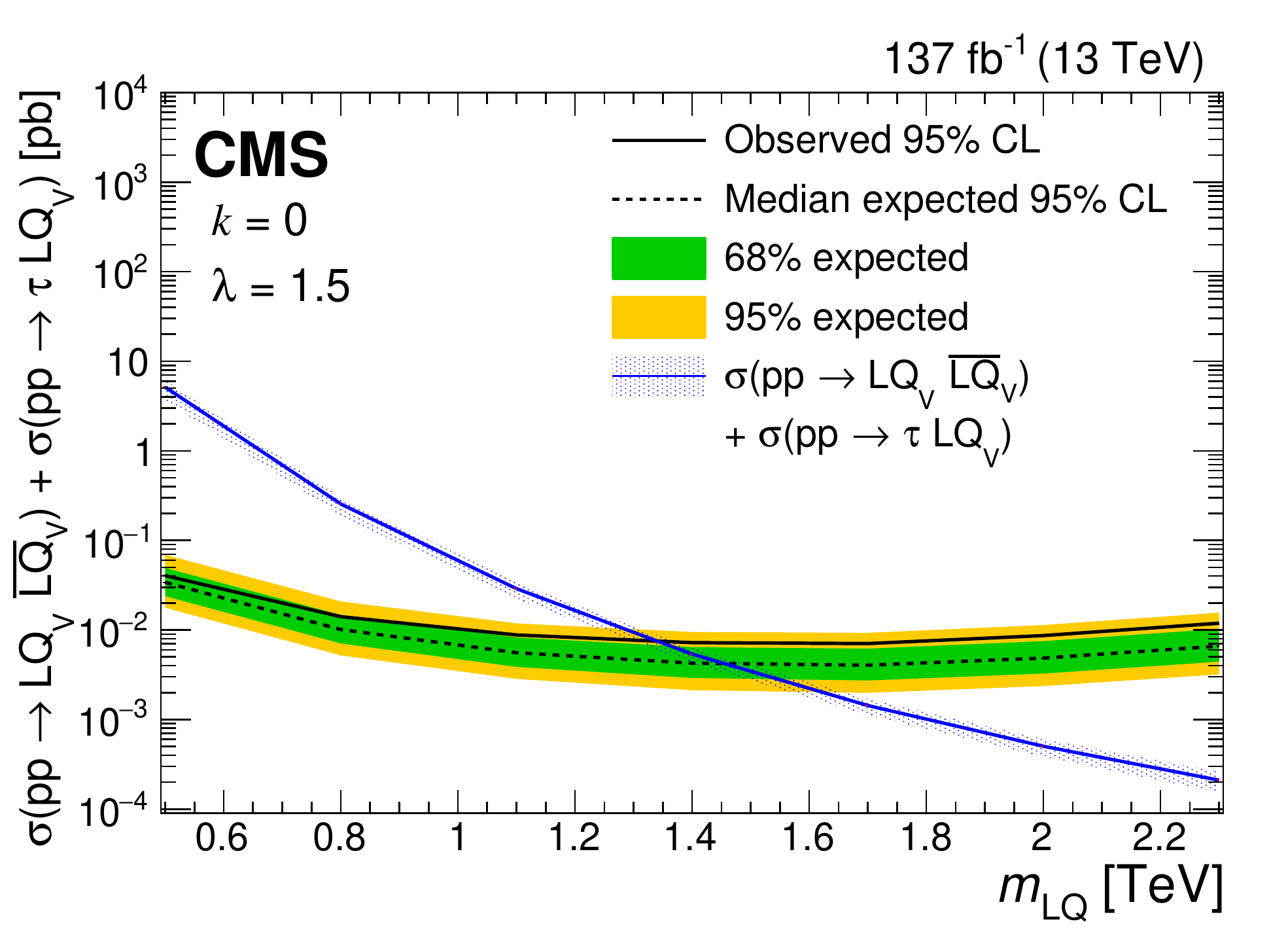}
\includegraphics[width=0.49\textwidth]{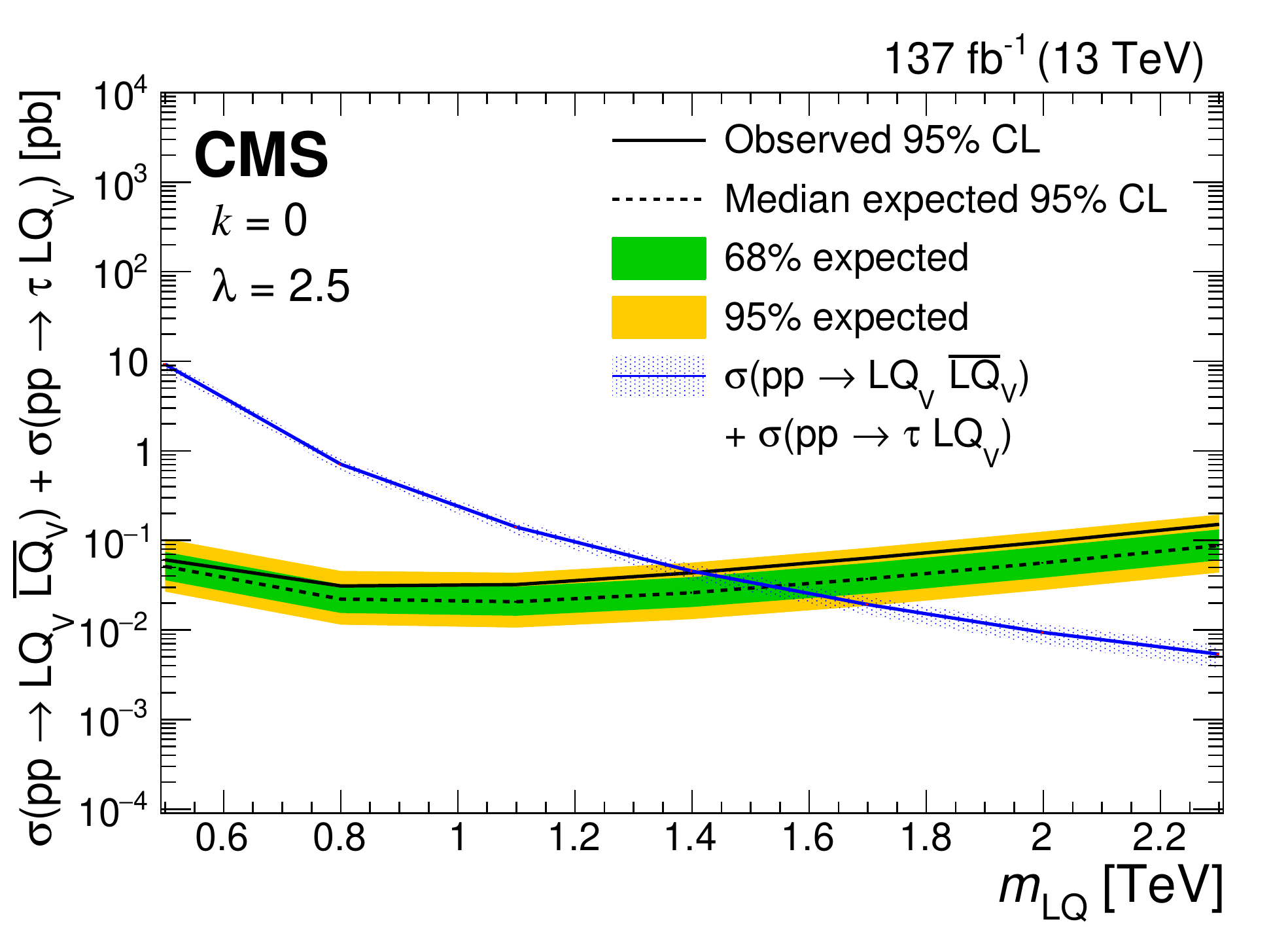}\\
\caption{The observed and expected (solid and dotted black lines) 95\% \CL upper limits on $\sigma(\Pp\Pp\to\LQV\ALQV)$ (upper), $\sigma(\Pp\Pp\to\tau\LQV)$ with $\lambda = 1.5$ and $2.5$ (middle left and right), and $\sigma(\Pp\Pp\to\LQV\ALQV)+\sigma(\Pp\Pp\to\tau\LQV)$ with $\lambda = 1.5$ and $2.5$ (lower left and right), as a function of the mass of the \LQV, with $k = 0$. The limits apply under the assumption of equal couplings for the \LQ decay to each of the two allowed lepton flavor pairings.
The bands represent the one- and two-standard deviation variations of the expected limit.
The solid blue curve indicates the theoretical predictions at LO. 
}\label{fig:LimLQvk0}
\end{figure*}

\begin{figure*}[!tbp]
\centering
\includegraphics[width=0.49\textwidth]{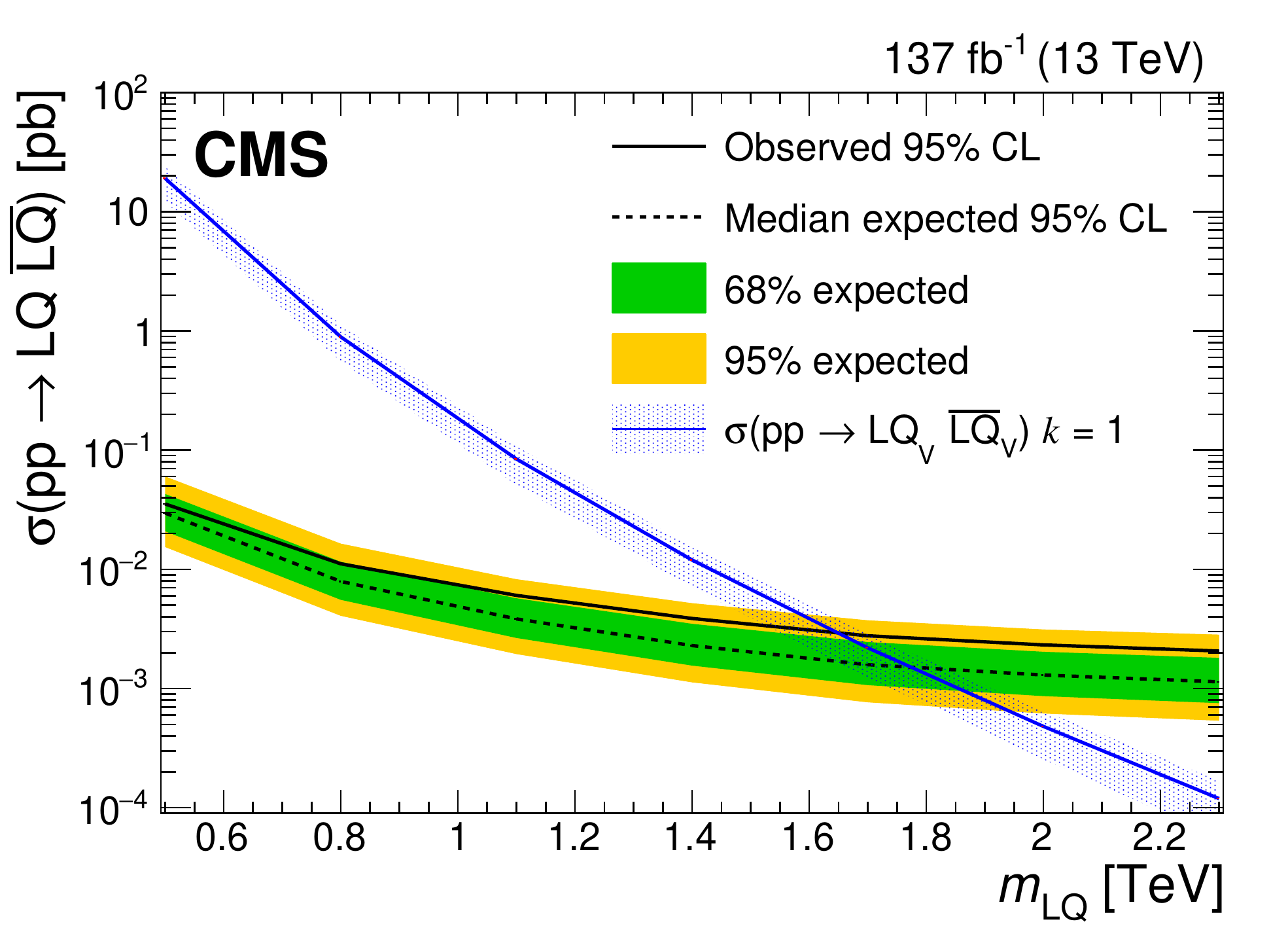}\\
\includegraphics[width=0.49\textwidth]{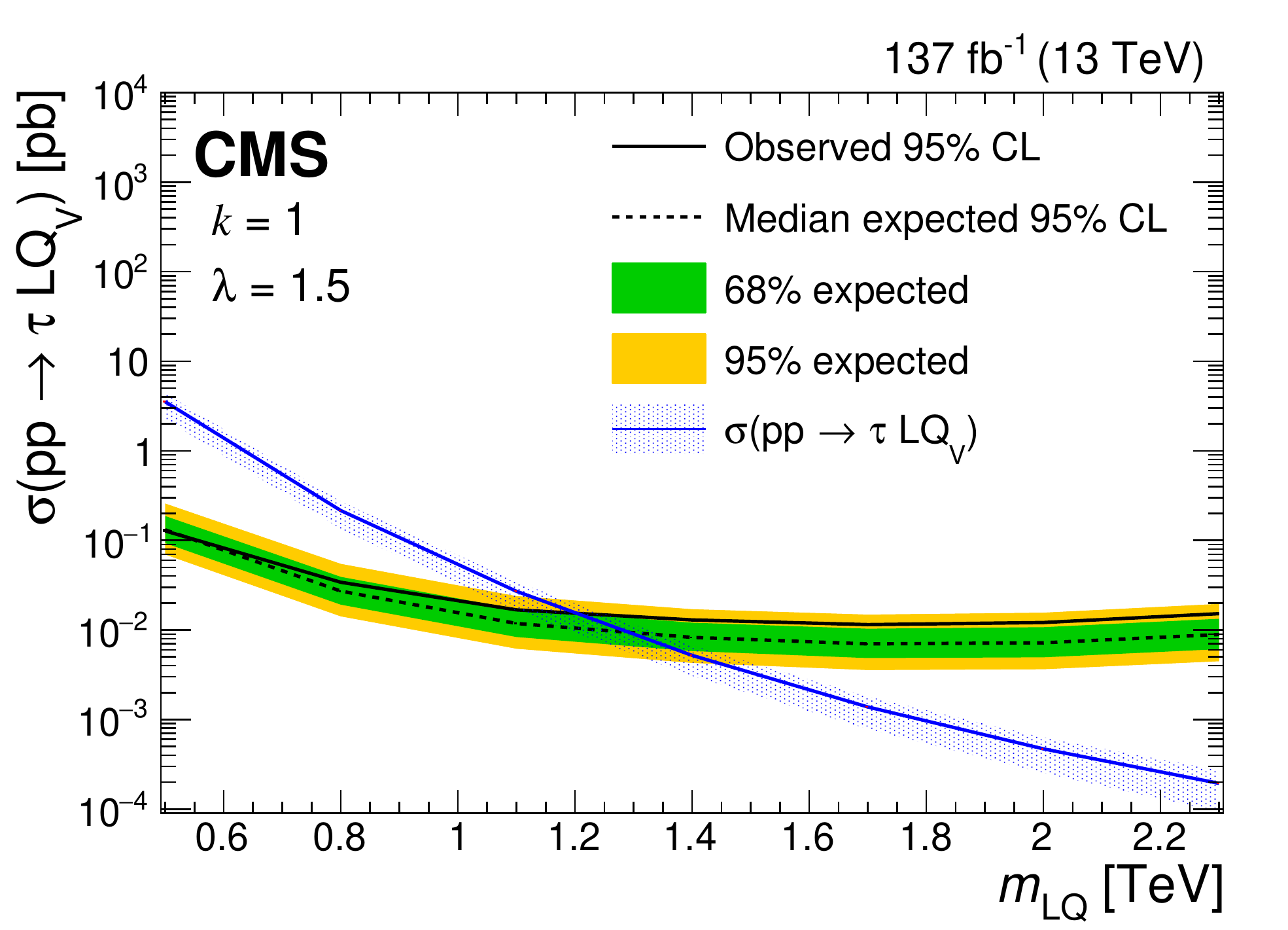}
\includegraphics[width=0.49\textwidth]{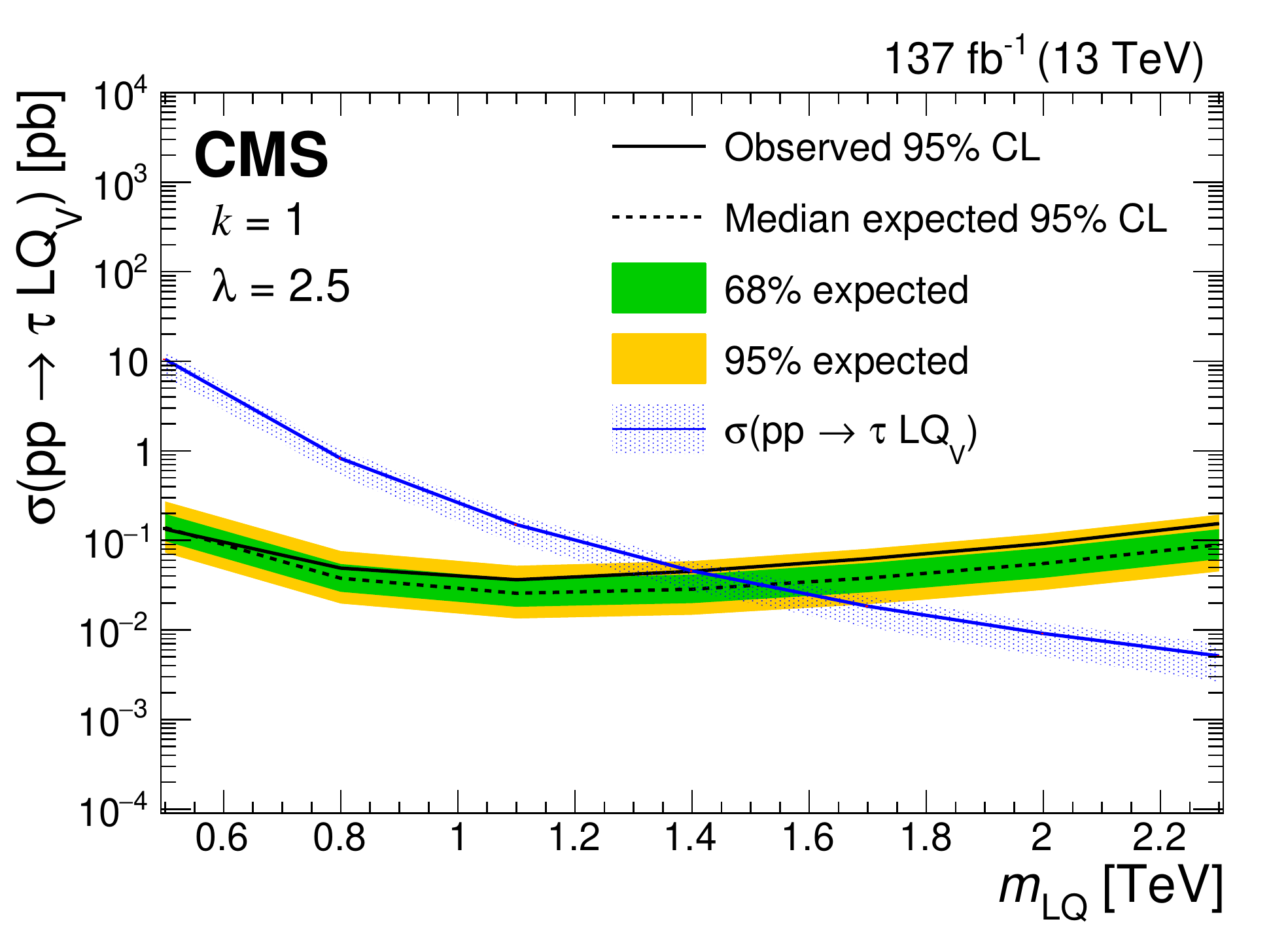}\\
\includegraphics[width=0.49\textwidth]{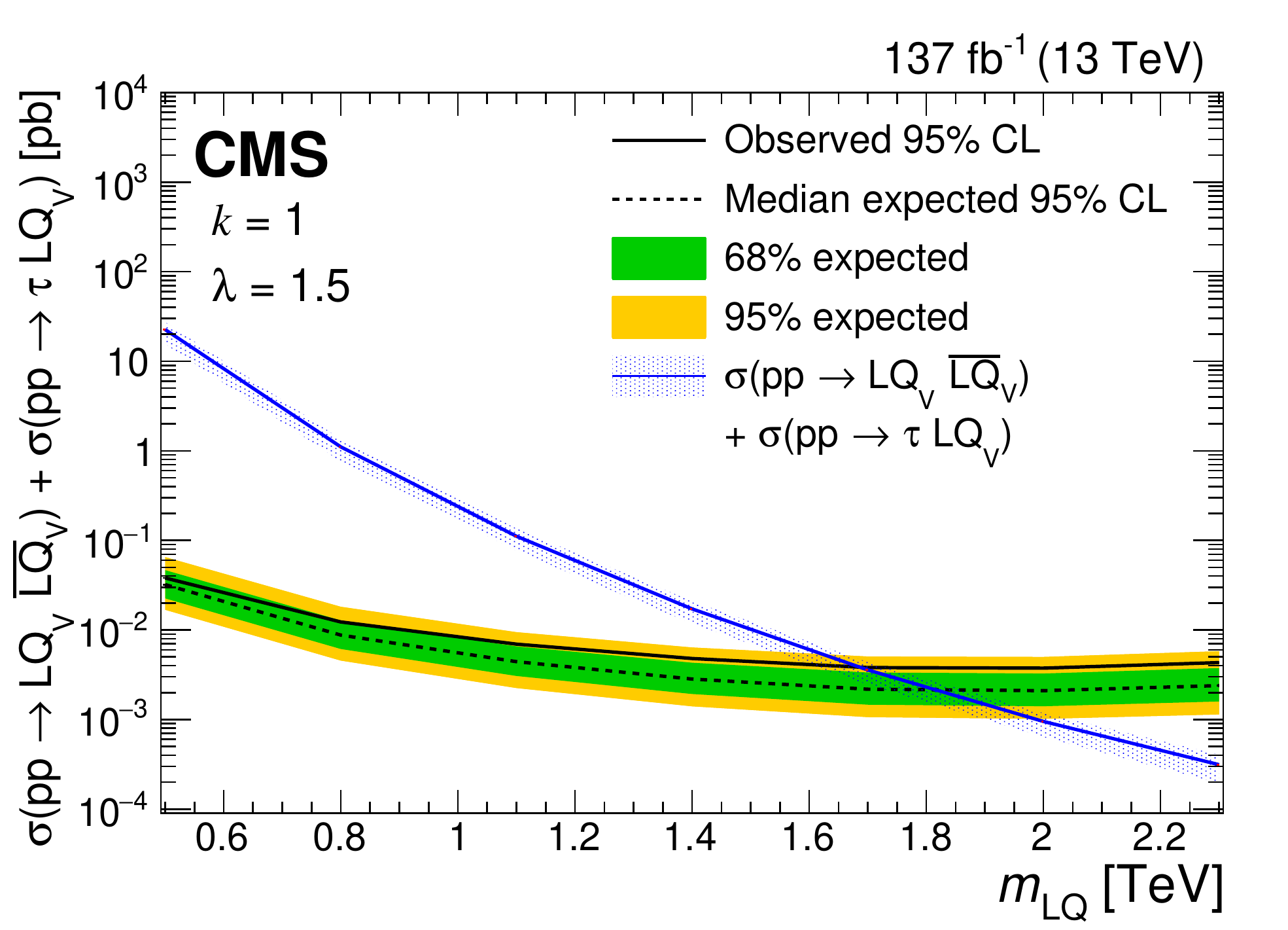}
\includegraphics[width=0.49\textwidth]{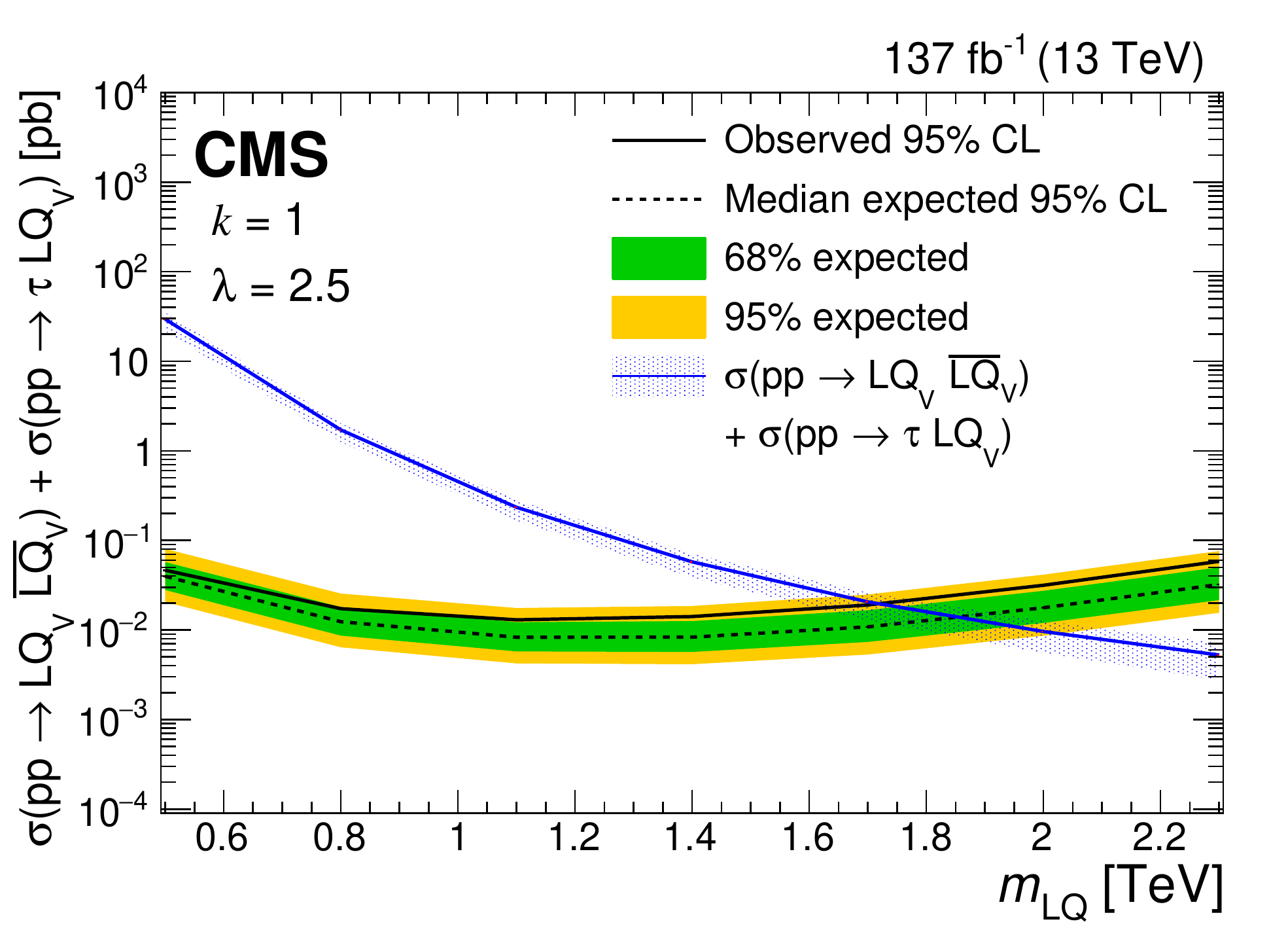}\\
\caption{The observed and expected (solid and dotted black lines) 95\% \CL upper limits on $\sigma(\Pp\Pp\to\LQV\ALQV)$ (upper), $\sigma(\Pp\Pp\to\tau\LQV)$ with $\lambda = 1.5$ and $2.5$ (middle left and right), and $\sigma(\Pp\Pp\to\LQV\ALQV)+\sigma(\Pp\Pp\to\tau\LQV)$ with $\lambda = 1.5$ and $2.5$ (lower left and right), as a function of the mass of the \LQV, with $k = 1$. The limits apply under the assumption of equal couplings for the \LQ decay to each of the two allowed lepton flavor pairings. 
The bands represent the one- and two-standard deviation variations of the expected limit.
The solid blue curve indicates the theoretical predictions at LO. 
}\label{fig:LimLQvk1}
\end{figure*}

The observed and expected upper limits on $\sigma(\Pp\Pp \to \LQ\ALQ)$, $\sigma(\Pp\Pp \to \ell\LQ$), and the case where both pair and single production mechanisms are considered simultaneously, $\sigma(\Pp\Pp \to \LQ\ALQ)+\sigma(\Pp\Pp \to \ell\LQ$), as a function of the
mass of the leptoquarks are shown in 
Figs.~\ref{fig:LimLQs}--\ref{fig:LimLQvk1},
where the leptoquarks are \LQS, \LQV $k = 0$, and \LQV $k = 1$, respectively. 
The uncertainty in the production cross section shown in these figures is given by the sum in quadrature of contributions arising from the PDFs and the renormalization and factorization scales. To estimate the latter, we consider the effects of multiplying these scales by factors of 0.5 and 2~\cite{Kalogeropoulos:2018cke,Catani:2003zt,Cacciari:2003fi}.
For single \LQ production, the limits are shown for fixed values of $\lambda = 1.5$ and 2.5.
Only values of $\lambda$ less than 2.5 are considered, since higher values are excluded by constraints from electroweak precision measurements~\cite{Dorsner:2016wpm}.
The bands represent the one- and two-standard deviation variations of the expected limit.
The solid blue curve indicates the theoretical prediction of $\sigma(\Pp\Pp \to \LQ\ALQ)$ and $\sigma(\Pp\Pp \to \ell\LQ)$, calculated at LO except for the pair production of \LQS, computed using NLO quantum chromodynamics corrections~\cite{Kramer:2004df} and the model implementation in Ref.~\cite{Dorsner:2018ynv}.
The intersection of the blue and the solid (dotted) black lines determines the observed (expected) lower limit on the \LQ mass. 
Table~\ref{tab:massLimits} summarizes the observed and expected lower limits on the LQ mass inferred from 
Figs.~\ref{fig:LimLQs}--\ref{fig:LimLQvk1} for the three cases, \LQS, \LQV $k = 0$, and \LQV $k = 1$.
The observed limits are, respectively, 0.98--1.02, 1.34--1.41, and 1.69--1.73\TeV for values of $\lambda$ between 1.5 and 2.5, based on the simultaneous search for single and pair production. The table also reports exclusion limits for the separate searches for single and pair production.

\begin{table*}[!ht]
\centering
 \topcaption{Lower limits on the mass in\TeV of the leptoquarks \LQS, \LQV $k = 0$, and \LQV $k = 1$, based on the pair- and single-production mechanisms taken either separately or together. These lower limits are derived from the intersection of the observed 95\% CL upper limits on the signal cross section and the signal cross section in Figs.~\ref{fig:LimLQs}--\ref{fig:LimLQvk1}. The results of the searches that depend on the $\lambda$ parameter are given for values of 1.5 and 2.5. The expected limits are given in parentheses.}
\label{tab:massLimits}
\begin{tabular}{lcccccc}
\hline
 & \multicolumn{2}{c}{\LQS (TeV)} & \multicolumn{2}{c}{\LQV $k = 0$ (TeV)} & \multicolumn{2}{c}{\LQV $k = 1$ (TeV)} \\ 
\hline 
Pair 
 & \multicolumn{2}{c}{0.95 (1.03)} & \multicolumn{2}{c}{1.29 (1.39)} & \multicolumn{2}{c}{1.65 (1.77)} \\
 & \multicolumn{1}{c}{$\lambda = 1.5$} & \multicolumn{1}{c}{2.5} & \multicolumn{1}{c}{1.5} & \multicolumn{1}{c}{2.5} & \multicolumn{1}{c}{1.5} & \multicolumn{1}{c}{2.5}\\ 
  
Single 
 & \multicolumn{1}{c}{0.55 (0.56)} & \multicolumn{1}{c}{0.75 (0.81)} & \multicolumn{1}{c}{1.03 (1.12)} & \multicolumn{1}{c}{1.25 (1.35)} & \multicolumn{1}{c}{1.20 (1.29)} & \multicolumn{1}{c}{1.41 (1.53)} \\

Pair+Single  
 & \multicolumn{1}{c}{0.98 (1.06)} & \multicolumn{1}{c}{1.02 (1.10)} & \multicolumn{1}{c}{1.34 (1.46)} & \multicolumn{1}{c}{1.41 (1.54)} & \multicolumn{1}{c}{1.69 (1.81)} & \multicolumn{1}{c}{1.73 (1.87)} \\
\hline 
\end{tabular}
\end{table*}

The combination of the two production mechanisms extends the exclusion on the \LQ mass by 30--120\GeV depending on the type of \LQ.
These exclusions represent the most stringent limits to date on the existence of \LQS (\LQV) coupled to 
$\PQt\tau$, $\PQb\nu$ ($\PQt\nu$, $\PQb\tau$) under the assumption of equal couplings to the lepton-quark pairs.
Comparing the cases of $\lambda = 1.5$ and 2.5 in Figs.~\ref{fig:LimLQs}--\ref{fig:LimLQvk1}, one can see how the upper limits on the cross section increase at higher \LQ masses and $\lambda$ values, as a result of an increasing relative contribution of virtual \LQ states in single \LQ production, as discussed in Section~\ref{sec:Data and simulated samples}, which degrades the sensitivity of the search. 

\begin{figure*}[!tb]
\centering
\includegraphics[width=0.49\textwidth]{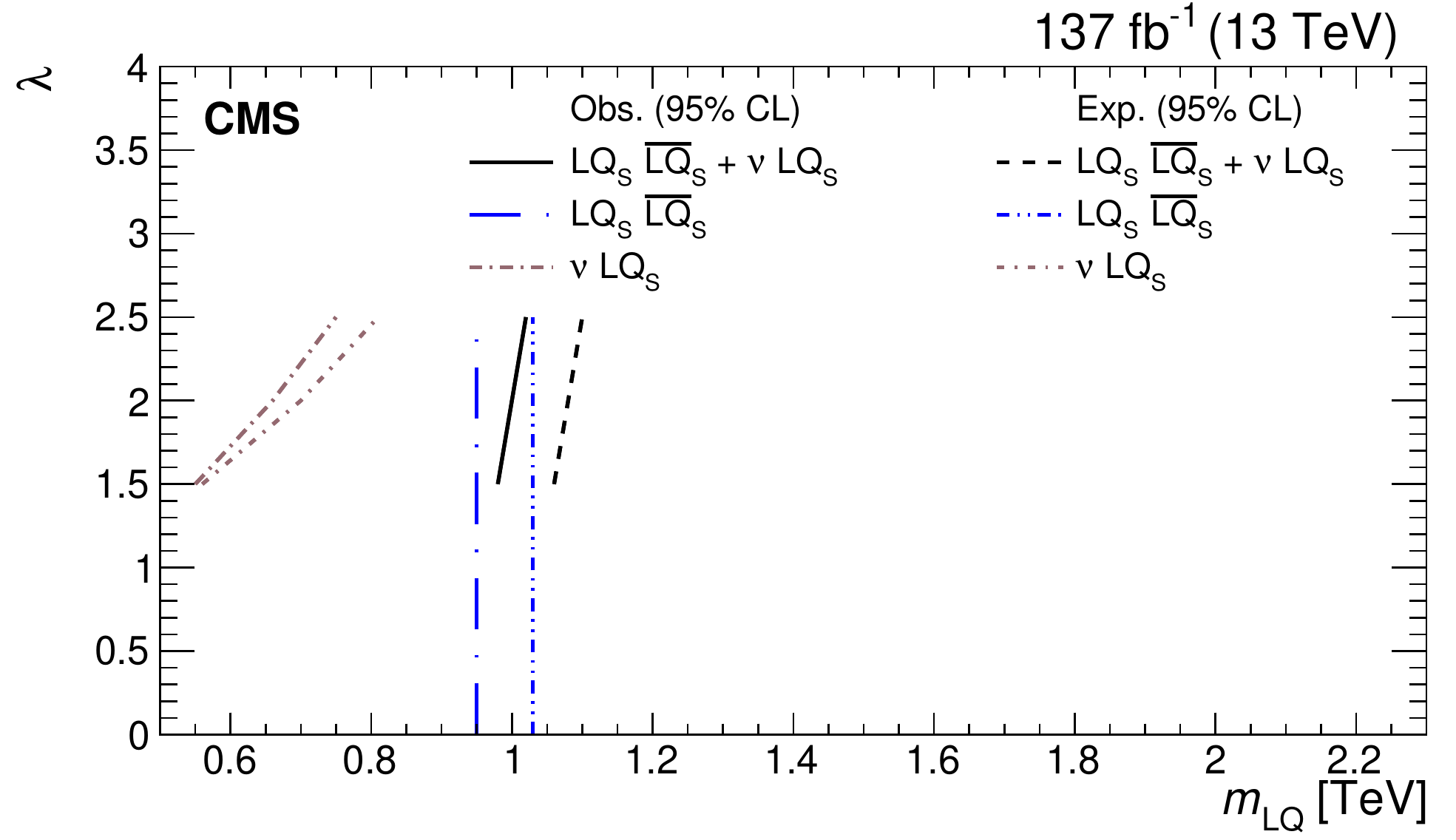}\\
\includegraphics[width=0.49\textwidth]{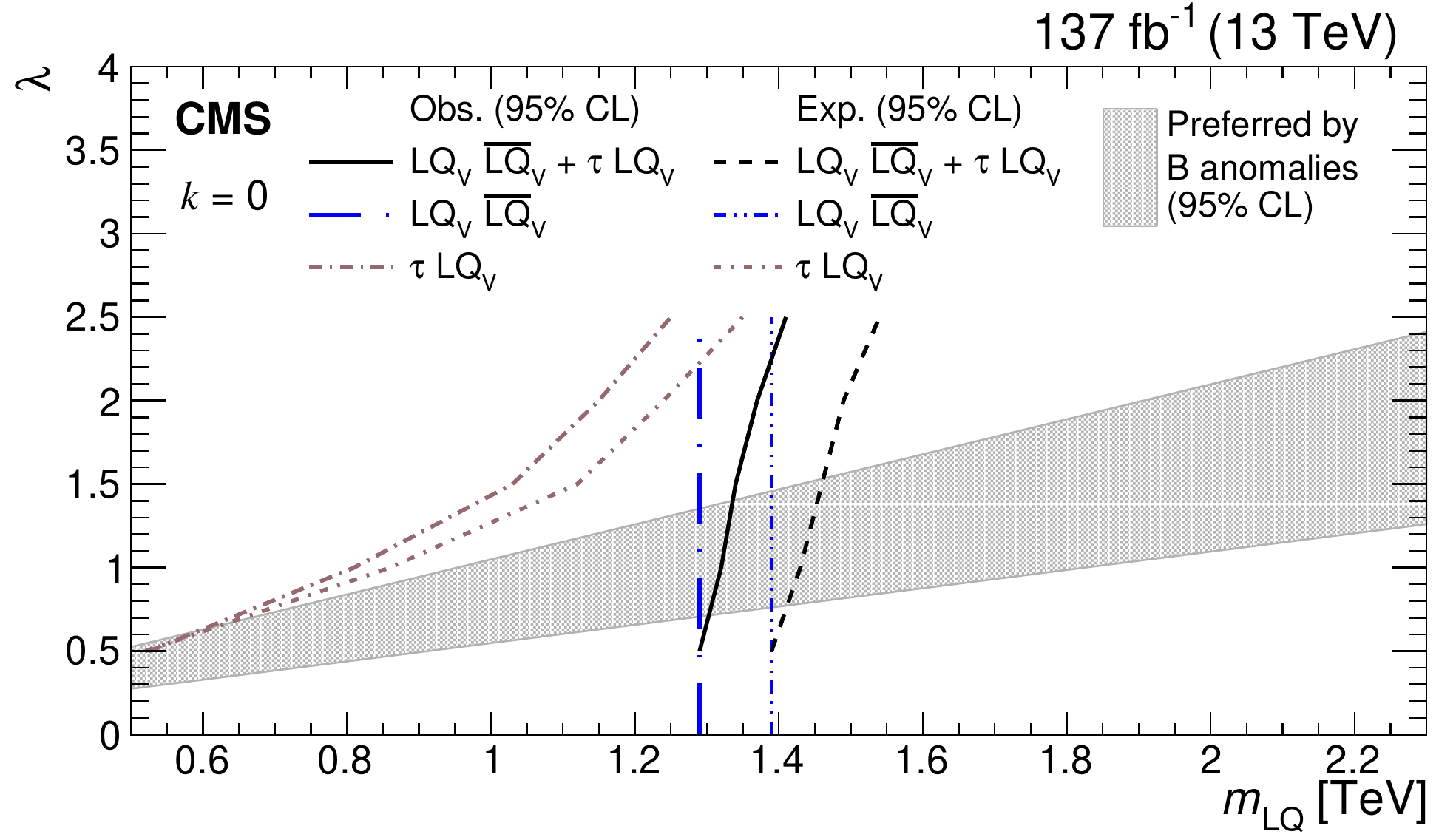}
\includegraphics[width=0.49\textwidth]{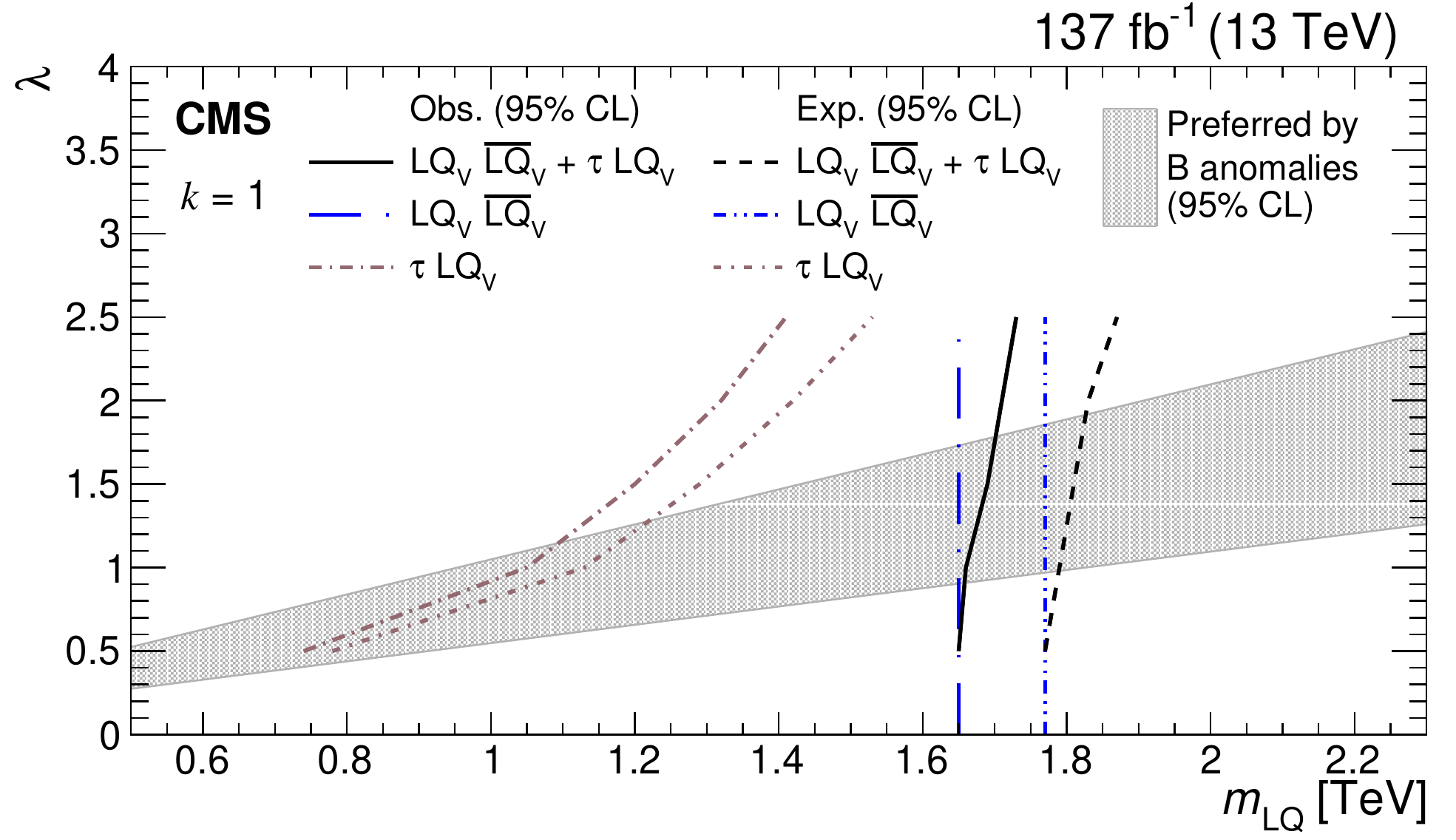}
\caption{The observed and expected 95\% \CL \LQ exclusion limits in the plane of the \LQ-lepton-quark coupling and the mass of the \LQ for single (brown lines) and pair (blue lines) production, and considering their sum (black lines). Regions to the left of the lines are excluded. The upper plot pertains to an \LQS with equal couplings to $\PQt\tau$ and $\PQb\nu$, while the lower plots are for an \LQV assuming $k = 0$ (left) and 1 (right) and equal couplings to $\PQt\nu$ and $\PQb\tau$. For \LQV, the gray area shows the band preferred (95\% \CL) by the \PB\ physics anomalies:
$\lambda = C m_{\mathrm{LQ}}$, where $C = \sqrt{0.7 \pm 0.2} \TeV^{-1}$ and $m_{\mathrm{LQ}}$ is expressed in \TeVns~\cite{Buttazzo:2017ixm}. 
}\label{fig:Lim2D}
\end{figure*}

Figure~\ref{fig:Lim2D} gives the observed and expected exclusion on the existence of leptoquarks in the $\lambda-m_{\LQ}$ plane,
for the single and pair production mechanisms and their combination.
For \LQV, the gray area shows the 95\% \CL band preferred by the \PB\ physics anomalies~\cite{Buttazzo:2017ixm},  
which is given by $\lambda = C m_{\mathrm{LQ}}$, where $C = \sqrt{0.7 \pm 0.2} \TeV^{-1}$ and $m_{\mathrm{LQ}}$ is expressed in \TeVns.
A relevant portion of this parameter space is excluded.

\section{Summary}\label{sec:summary}
A search for leptoquarks coupled to third-generation fermions, and produced in pairs and singly in association with a lepton, has been presented.
The leptoquark (\LQ) may couple to a top quark and a $\tau$ lepton ($\PQt\PGt$) or a bottom quark and a neutrino ($\PQb\PGn$, scalar \LQ) 
or else to $\PQt\PGn$ and $\PQb\PGt$ (vector \LQ), 
resulting in the $\PQt\PGt\PGn\PQb$ and $\PQt\PGt\PGn$ signatures. The channel in which both the top quark and the $\tau$ lepton decay hadronically is investigated, including the
case of a large \LQ-$\PQt$ mass splitting giving rise to a Lorentz-boosted top quark, whose decay daughters may not be resolved as individual jets. 
This particular signature has not been previously examined in searches for physics beyond the standard model. The data used corresponds to an integrated luminosity of $\runtwolumi\fbinv$ collected with the CMS detector at the CERN LHC in proton-proton collisions at $\sqrt{s} = 13\TeV$. 
The observations are found to be in agreement with the standard model predictions.
Exclusion limits are given in the plane of the \LQ-lepton-quark vertex coupling $\lambda$ and the \LQ mass for scalar and vector leptoquarks. 
The range of lower limits on the \LQ mass, at 95\% confidence level, is 0.98--1.73\TeV, depending on $\lambda$ and the leptoquark spin.
These results represent the most stringent limits to date on the existence of such \LQs for the case of equal couplings to the lepton-quark pairs.
They allow a relevant portion of the parameter space preferred by the B-physics anomalies in several models \cite{Alvarez:2018gxs,Buttazzo:2017ixm} to be excluded. 

\begin{acknowledgments}
  We congratulate our colleagues in the CERN accelerator departments for the excellent performance of the LHC and thank the technical and administrative staffs at CERN and at other CMS institutes for their contributions to the success of the CMS effort. In addition, we gratefully acknowledge the computing  centers and personnel of the Worldwide LHC Computing Grid for delivering so effectively the computing infrastructure essential to our analyses. Finally, we acknowledge the enduring support for the construction and operation of the LHC and the CMS detector provided by the following funding agencies: BMBWF and FWF (Austria); FNRS and FWO (Belgium); CNPq, CAPES, FAPERJ, FAPERGS, and FAPESP (Brazil); MES (Bulgaria); CERN; CAS, MoST, and NSFC (China); COLCIENCIAS (Colombia); MSES and CSF (Croatia); RIF (Cyprus); SENESCYT (Ecuador); MoER, ERC PUT and ERDF (Estonia); Academy of Finland, MEC, and HIP (Finland); CEA and CNRS/IN2P3 (France); BMBF, DFG, and HGF (Germany); GSRT (Greece); NKFIA (Hungary); DAE and DST (India); IPM (Iran); SFI (Ireland); INFN (Italy); MSIP and NRF (Republic of Korea); MES (Latvia); LAS (Lithuania); MOE and UM (Malaysia); BUAP, CINVESTAV, CONACYT, LNS, SEP, and UASLP-FAI (Mexico); MOS (Montenegro); MBIE (New Zealand); PAEC (Pakistan); MSHE and NSC (Poland); FCT (Portugal); JINR (Dubna); MON, RosAtom, RAS, RFBR, and NRC KI (Russia); MESTD (Serbia); SEIDI, CPAN, PCTI, and FEDER (Spain); MOSTR (Sri Lanka); Swiss Funding Agencies (Switzerland); MST (Taipei); ThEPCenter, IPST, STAR, and NSTDA (Thailand); TUBITAK and TAEK (Turkey); NASU (Ukraine); STFC (United Kingdom); DOE and NSF (USA).
  
  \hyphenation{Rachada-pisek} Individuals have received support from the Marie-Curie prog ram and the European Research Council and Horizon 2020 Grant, contract Nos.\ 675440, 724704, 752730, and 765710 (European Union); the Leventis Foundation; the A.P.\ Sloan Foundation; the Alexander von Humboldt Foundation; the Belgian Federal Science Policy Office; the Fonds pour la Formation \`a la Recherche dans l'Industrie et dans l'Agriculture (FRIA-Belgium); the Agentschap voor Innovatie door Wetenschap en Technologie (IWT-Belgium); the F.R.S.-FNRS and FWO (Belgium) under the ``Excellence of Science -- EOS" -- be.h project n.\ 30820817; the Beijing Municipal Science \& Technology Commission, No. Z191100007219010; the Ministry of Education, Youth and Sports (MEYS) of the Czech Republic; the Deutsche Forschungsgemeinschaft (DFG) under Germany's Excellence Strategy -- EXC 2121 ``Quantum Universe" -- 390833306; the Lend\"ulet (``Momentum") Prog ram and the J\'anos Bolyai Research Scholarship of the Hungarian Academy of Sciences, the New National Excellence Program \'UNKP, the NKFIA research grants 123842, 123959, 124845, 124850, 125105, 128713, 128786, and 129058 (Hungary); the Council of Science and Industrial Research, India; the HOMING PLUS prog ram of the Foundation for Polish Science, cofinanced from European Union, Regional Development Fund, the Mobility Plus prog ram of the Ministry of Science and Higher Education, the National Science Center (Poland), contracts Harmonia 2014/14/M/ST2/00428, Opus 2014/13/B/ST2/02543, 2014/15/B/ST2/03998, and 2015/19/B/ST2/02861, Sonata-bis 2012/07/E/ST2/01406; the National Priorities Research Program by Qatar National Research Fund; the Ministry of Science and Higher Education, project no. 0723-2020-0041 (Russia); the Tomsk Polytechnic University Competitiveness Enhancement Program; the Programa Estatal de Fomento de la Investigaci{\'o}n Cient{\'i}fica y T{\'e}cnica de Excelencia Mar\'{\i}a de Maeztu, grant MDM-2015-0509 and the Programa Severo Ochoa del Principado de Asturias; the Thalis and Aristeia prog rams cofinanced by EU-ESF and the Greek NSRF; the Rachadapisek Sompot Fund for Postdoctoral Fellowship, Chulalongkorn University and the Chulalongkorn Academic into Its 2nd Century Project Advancement Project (Thailand); the Kavli Foundation; the Nvidia Corporation; the SuperMicro Corporation; the Welch Foundation, contract C-1845; and the Weston Havens Foundation (USA).
\end{acknowledgments}

\clearpage
\bibliography{auto_generated}
\cleardoublepage \appendix\section{The CMS Collaboration \label{app:collab}}\begin{sloppypar}\hyphenpenalty=5000\widowpenalty=500\clubpenalty=5000\vskip\cmsinstskip
\textbf{Yerevan Physics Institute, Yerevan, Armenia}\\*[0pt]
A.M.~Sirunyan$^{\textrm{\dag}}$, A.~Tumasyan
\vskip\cmsinstskip
\textbf{Institut f\"{u}r Hochenergiephysik, Wien, Austria}\\*[0pt]
W.~Adam, T.~Bergauer, M.~Dragicevic, J.~Er\"{o}, A.~Escalante~Del~Valle, R.~Fr\"{u}hwirth\cmsAuthorMark{1}, M.~Jeitler\cmsAuthorMark{1}, N.~Krammer, L.~Lechner, D.~Liko, I.~Mikulec, F.M.~Pitters, N.~Rad, J.~Schieck\cmsAuthorMark{1}, R.~Sch\"{o}fbeck, M.~Spanring, S.~Templ, W.~Waltenberger, C.-E.~Wulz\cmsAuthorMark{1}, M.~Zarucki
% ...
\vskip\cmsinstskip
\dag: Deceased\\
1:  Also at Vienna University of Technology, Vienna, Austria\\
% ...
\end{sloppypar}
\end{document}